\documentclass[12pt,preprint]{aastex}
\newcommand{\be}{\begin{eqnarray}}
\newcommand{\ee}{\end{eqnarray}}
\newcommand{\Gl}[1]{Eq.~(\ref{#1})}
\newcommand{\Fig}[1]{Fig.~\ref{#1}}

\def\gr{$\gamma$-ray }

\shorttitle{Cosmic-ray propagation properties}
\shortauthors{B\"usching et al.}

\begin{document}

\title{Cosmic-ray propagation properties for an origin in SNRs}

\author{I. B\"usching}
\affil{Ruhr-Universit\"at Bochum, 44780 Bochum, Germany}

\author{A. Kopp}
\affil{MPI f\"ur Sonnensystemforschung, 37191 Katleburg Lindau, Germany}

\author{M. Pohl}
\affil{Department of Physics and Astronomy, Iowa State University, Ames, IA 50011-3160, USA}

\author{R. Schlickeiser}
\affil{Ruhr University Bochum, 44780 Bochum, Germany}
\author{C. Perrot}
\affil{W.W. Hansen Experimental Physics Laboratory, Stanford University, Stanford, CA
94305, USA}
\author{I.  Grenier}
\affil{Universit\'e Paris VII \& CEA/Saclay, Service d'Astrophysique, 91191 Gif-sur-Yvette, France}
\begin{abstract} 
 
We  have studied the impact of cosmic-ray acceleration in SNR
on the spectra of cosmic-ray nuclei in the Galaxy using a series expansion 
of the propagation equation, 
which allows us to use analytical solutions for part of the problem and
an efficient numerical treatment of the remaining equations and thus
accurately describes the cosmic-ray propagation on small scales around their sources in
three spatial dimensions and time.
We found strong variations of the cosmic-ray nuclei flux by typically 20\% with 
occasional spikes of much higher amplitude,
but only minor changes in the spectral distribution.  
The locally measured spectra of primary cosmic rays fit well into the  obtained range
of possible spectra. 
We further showed that the spectra of the secondary element Boron 
show almost no variations, so that 
the above  findings also imply significant fluctuations of
the Boron-to-Carbon ratio.
Therefore the commonly used method of determining CR propagation parameters 
by fitting secondary-to-primary ratios appears flawed on account of the
variations that these ratios would show throughout the Galaxy.
\end{abstract}

\keywords{cosmic ray propagation, cosmic ray origin}
\section{Introduction}
 The origin of cosmic-rays (CR) is one of the major problems in modern astrophysics.
In that quest two possible strategies may be followed:
\begin{itemize}
\item one may search for high-energy emission from the source regions themselves, for 
the CR density must be very high therein.
Particle acceleration at supernova remnant (SNR) shock waves is regarded as
the most probable mechanism for providing Galactic cosmic rays at energies
below $10^{15}$ eV on account of the power requirements \citep{be87}. 
The recent detections of non-thermal X-ray synchrotron radiation
from the four supernova remnants SN1006 \citep{koyama95}, RX 
J1713.7-3946 \citep{koyama97,sla99}, Cas A 
\citep{allen97,go01}, and RCW86 \citep{ba00,bor01,rho02}  
support the hypothesis that at least CR electrons are 
accelerated predominantly in SNR. The evidence for CR nucleon acceleration in SNR is at best
controversial \citep{bkv02,reimer02}.
\item one may also construct models for the diffusive propagation of CRs in the Galaxy
and compare their predictions with observed CR spectra at earth
\citep[e.g.][]{strong98,maurin01,taillet03}. The relation between primary CRs, which have 
been accelerated somewhere in the Galaxy, and secondary CRs, which are produced in 
interactions of CRs with ambient gas, then allows to determine the propagation parameters
such as the diffusion coefficient, the size of the diffusion region, and others.
Nearly all published studies of that kind assume steady-state conditions.
\end{itemize}

 Already in 1969, \citet{lingenfelter69} found the CR energy density at earth
to vary due to nearby SN, given the CR are accelerated at these sites.
A few years ago \citet{pohl98} showed for CR electrons that in case of SN origin  
the CR electron spectrum would vary in space and time throughout the Galaxy.
The purpose of this paper is to present a time-dependent propagation model for
CR nucleons. In particular we aim to address the following questions:
\begin{itemize}
\item would a SNR origin of CR nucleons also lead to significant fluctuations of the CR
density in the Galaxy, which then would modify secondary-to-primary ratios from their 
steady-state values? If that was the case, we would have to re-think the way we infer
CR propagation parameters from their locally observed spectra.
\item are there any signatures in the CR distribution in the Galaxy, that might permit to
infer a SNR origin of CR nucleons on the grounds of locally observed CR spectra and the
diffuse Galactic gamma-ray emission?
\end{itemize}

For that purpose we have developed a method
to numerically solve the propagation equation based on a series expansion, 
which allows us to use analytical solutions for part of the problem and
an efficient numerical treatment of the remaining equations. As described in detail
in section \ref{sec:computations} this technique allows 
us to accurately follow the CR propagation on small scales around their sources in
three spatial dimensions and time.
In section \ref{sec:signatures1} we discuss
the results of these calculations with respect to the CR density distribution
and in section \ref{sec:disskuss} with respect to CR spectra, respectively.

\section{The Model}
\label{sec:computations}
\subsection{The basic equations}
\begin{figure}
\includegraphics{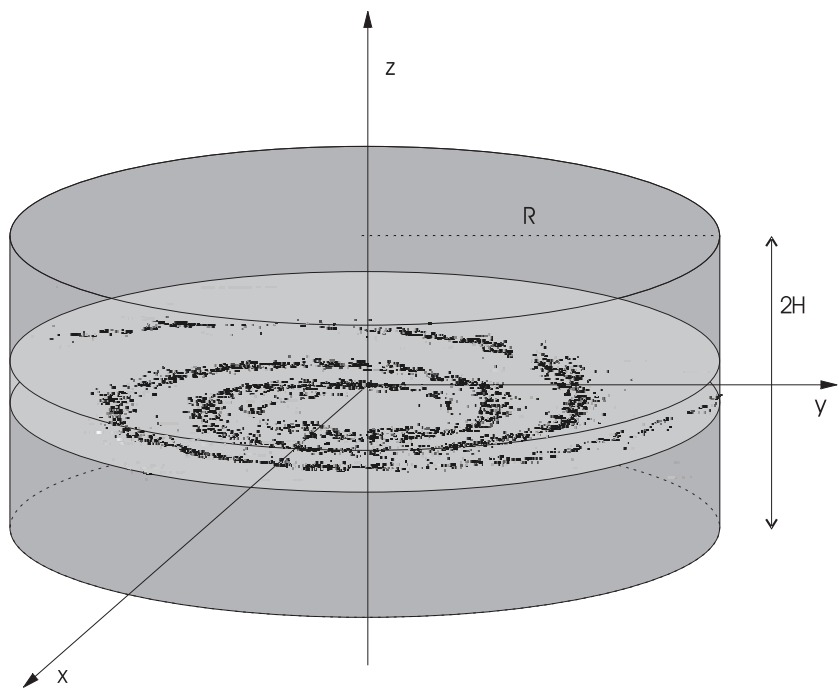}
\caption{
Sketch of the geometry used in our calculations. 
We assume that the Galactic disk with radius $R$  is filled
with interstellar gas and the CR sources. The density of the interstellar gas 
decreases quasi-exponentially in the halo.
The height of the entire diffusion zone is $2H$.
}
\label{fig:gage}
\end{figure}

 Our investigations are performed in the
framework of  a diffusion model of CR propagation, so we use a continuity
equation for the differential CR number density, $N$:
\be
\frac{\partial N}{\partial t}-S = \nabla \left(k \nabla N\right) 
 - \Omega v \sigma N.
\label{gle:mod}
\ee
with particle speed $v$ and total spallation cross section $\sigma$.  
We consider  the diffusion zone to be a
disk of radius $R=20$\,kpc  \citep[cf.][]{webber92} and height $2H$ (see \Fig{fig:gage}), where
$H$ is determined below.
The density of the interstellar gas is assumed to decrease
with distance from the galactic plane, $z$, with a profile
\be
\Omega(z)&=& \frac{n_0}{{\rm cosh}\left(z\,h_g\right)} \, .
\label{rech:gasd}
\ee
The parameters 
$n_0$ = 1.24\,cm$^{-3}$ and  $h_g$\,=\,30\,kpc$^{-1}$  corresponding 
to  a column density of 
$\sim 6.2\cdot 10^{20}\,$cm$^{-2}$ 
are chosen to be consistent with 
%as it is used  
the calculations by, e.g., \citet{webber92} or
  \citet{berezinskii90}.
The chemical composition of the ISM, which 
consists mainly of hydrogen and
some  heavier elements, is taken into account by multiplying 
the gas  density \Gl{rech:gasd} by a factor of 1.3, as it is done  by 
 \citet{mannheim94}  for the calculation of CR-induced pion production in the ISM. 

The diffusion coefficient, $k$, depends on the particle 
rigidity $\zeta$\,=\,$p/q$ with particle momentum $p$ and charge $q$. The form
\be
k&=&  \left\{  { k_0 \left(\frac{\zeta}{\zeta _0}\right)^{0.6} \,\,\,\,\,\qquad {\rm for\,\, } 
\zeta \geq \zeta _0 \atop
 k_0\left(\frac{\zeta}{\zeta _0}\right)^{-0.48} \qquad {\rm for\,\, } \zeta <\zeta _0 
} 
\right.
\label{rech:k12}
\ee 
with $\zeta _0 = 4$\,GV/c
is chosen to reproduce the observed Boron-to-Carbon ratio. 

As we consider discrete, point-like sources,  the source term $S$ in \Gl{gle:mod},
is  in fact the sum of the contributions of many
source, each of which has the same temporal and spectral form. For the time dependence
we assume a linear growth with an
exponential cut-off, with the source spectra being power laws  with index $s$ in particle 
rigidity $\zeta$. 
\begin{eqnarray}
q_j \left(\zeta,t \right)= \hat q_j\cdot \left(t-t_j\right){\rm exp}\left(-\frac{t-t_j}{20 \, 
{\rm kyr}}\right) 
\Theta\left(t-t_j\right) \cdot \left(\frac{\zeta}{\zeta_0}\right)^{-s}.
\label{rech:sstrength}
\end{eqnarray}

We assume the SN explosions to be stochastic events. Their effect on the CR spectra can thus be studied
by the method of Monte-Carlo simulations, so many possible CR spectra are calculated using 
randomly chosen sets of CR sources.
We used the rejection method \citep{press93} to compute 
the quantities $r_j,\varphi_j,z_j$ for the location, 
$t_j$  for the ignition time, and $\hat q_j$ for the source strength. The position of the CR sources
in azimuth, $\varphi_j$, is uniformly distributed, whereas for the radial 
distribution we use the form suggested by \citet{case96}, and for
the vertical distribution we use the same profile as for the density of interstellar gas. Then
\be
\tilde P_S(r,z)=\frac{1}{{\rm cosh}(z\,h_g)}
\left(\frac{b \,r}{a\, r_s}\right)^{a}\cdot {\rm exp}\left(\frac{a \,r_s-b \,r}{r_s}\right).
\label{rech:sourcedis}
\ee 
with parameters $a=1.69$, $b=3.33$, $r_s=8.5$ as taken from the paper of \citet{case96}.

The ignition time, $t_j$, and the source strength, $\hat q_j$, are uniformly distributed on the intervals $[t_{\rm start};t_{\rm end}]$ and $[0;1]$, respectively.
We have also performed simulations using a detailed model of the spatial and temporal
evolution of the nearby star-forming region Gould's belt \citep{perrot03}.

\begin{figure}
\begin{tabular}{l}
\begin{minipage}{\columnwidth}
\includegraphics[width=0.7\columnwidth]{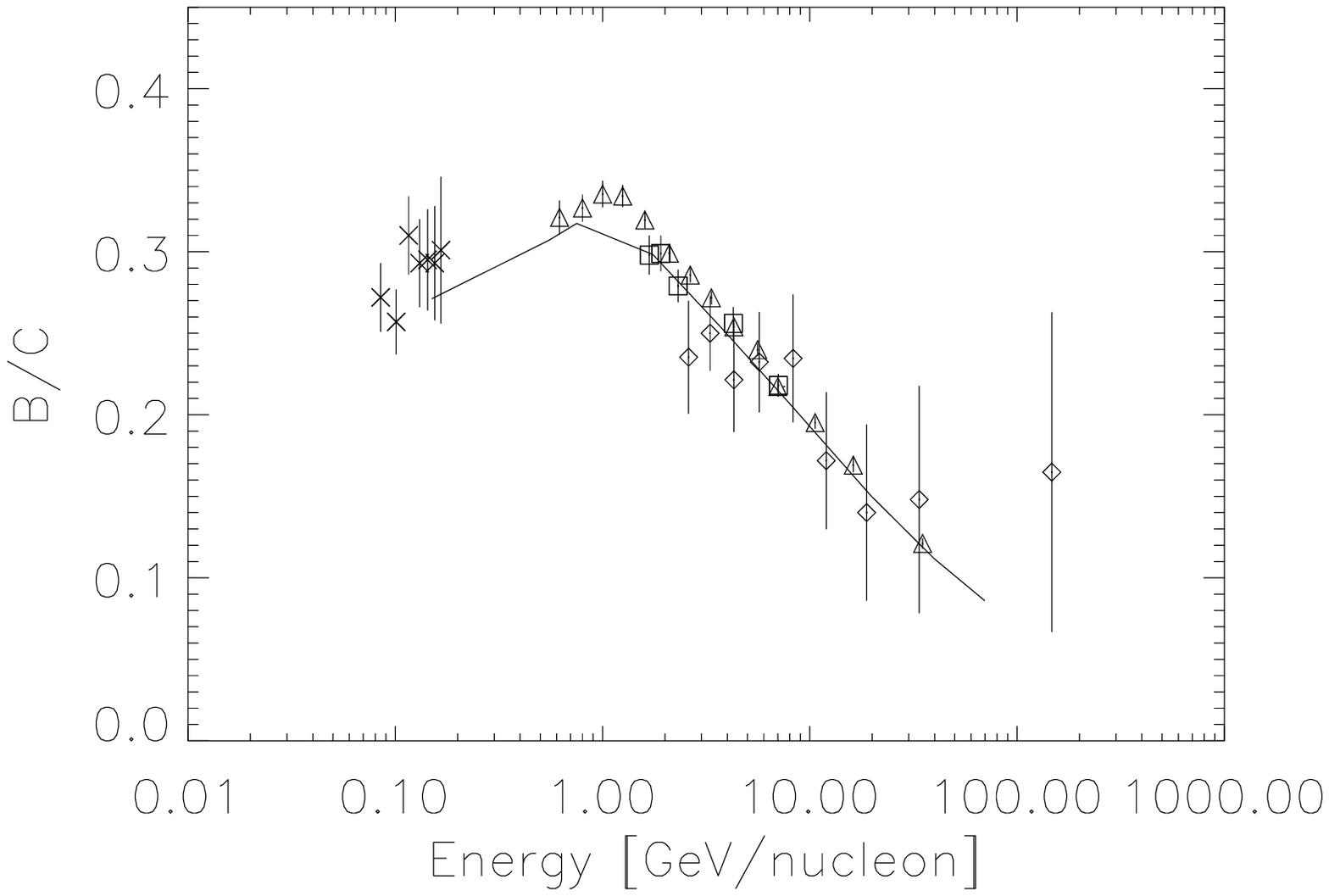}
 \end{minipage}
\\
      \begin{minipage}{\columnwidth}
\includegraphics[width=0.7\columnwidth]{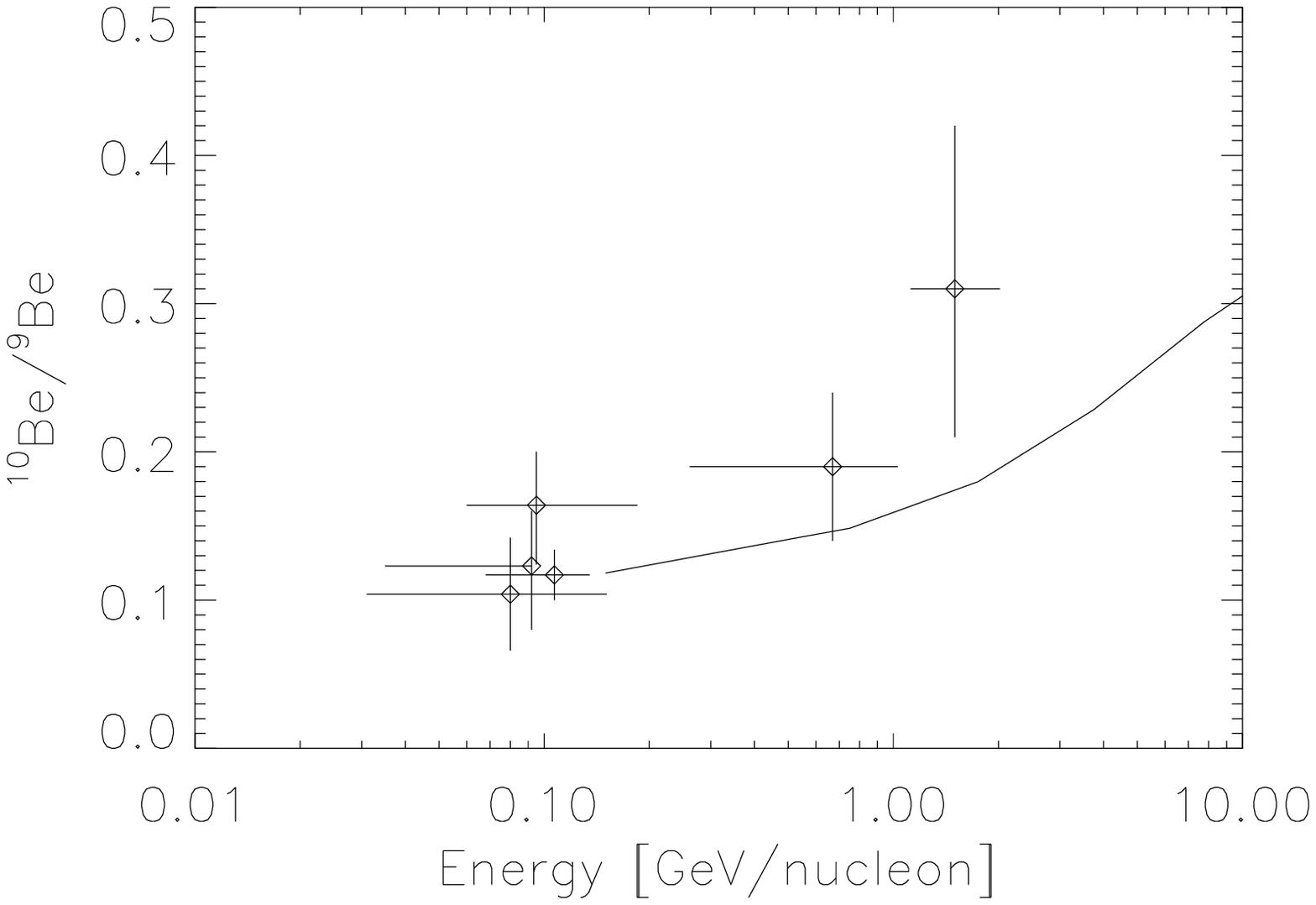}
\vspace{.0pc}
      \end{minipage}
\\
\end{tabular}
\caption{
Best fits of Boron-to-Carbon data (upper panel) and $^{10}$Be/$^9$Be data (lower panel), 
compared to data by \citet{engelmann90}  ($\triangle$), \citet{dwyer87} ($\Diamond$), 
\citet{krombel88} ($\Box$), and \citet{orth78} ($\times$)
for  the Boron-to-Carbon ratio and from 
\citet{connell98}, \citet{garcia-munoz81}, 
\citet{hams01}, \citet{lukasiak94}, 
 \citet{denolfo01}, and \citet{wiedenbeck80}
for the $^{10}$Be to
 $^{9}$Be ratio,
respectively.
}
\label{fig:fits}
\end{figure}

This leaves  $k_0$, the halo height $H$ and 
the source spectral index $s$ as free parameters, which 
we determined by fitting the Boron-to-Carbon ratio and the survival
fraction of $^{10}$Be calculated for the steady-state case to measured data.
Solar modulation is taken into account
using the force field approximation \citep{gleeson68} with modulation parameter $\Phi\,=\,500\,$MV.
In the steady-state case the best-fit parameters are $k_0=  0.26\,$kpc$^2$Myr$^{-1}$,
$H$ = 2\,kpc, and $s$ = 2.1, as shown in \Fig{fig:fits}. 

\subsection{The method of computation}
To determine the influence of the discrete nature of SNR as CR sources on
the CR distribution and spectra, one has to solve the CR propagation equation \Gl{gle:mod}
with high resolution, both in space and time. 
With today's computers, however, it is not possible to calculate the CR density
with a spatial resolution  sufficient to investigate point-like 
sources in an adequate way. 
{One should note that the grid size in finite-difference
algorithms, e.g. those in the widely used code  {\sc Galprop}
 \citep{strong01}, in principle has to be much finer than the size of the
presumed sources, for
one has to deal with large gradients in the CR flux on small scales. Thus, the accuracy of
representation of the CR density on a fixed grid is highly questionable. One way out of this quandary
may be the implementation of adaptive mesh-refinement techniques in the finite-difference
algorithms. The way we followed is to use an analytical ansatz 
that breaks down the  problem of solving the propagation equation \Gl{gle:mod}
into many small tasks, that are easily solved on pc-style hardware.}
The series ansatz to solve \Gl{gle:mod} will be in detail described
in the next section.
We also developed a method 
to spread the task of solving the propagation 
equation onto many computers, so the spatial resolution obtained is only
limited by the number of computers available.

\subsection{The Ansatz}
Considering the cylindrical geometry of our Galaxy (with radius $R$ and 
height $2H$, see \Fig{fig:gage}), we
assume the gas distribution $\Omega$ 
and  diffusion coefficient $k$ to be independent of $r$ and $\varphi$.
Then the CR transport equation \Gl{gle:mod} can be written in cylindrical coordinates:
\be
\frac{\partial N}{\partial t}-S &=& k(p) \left\{\frac{1}{r}\frac{\partial N}{\partial r}+
\frac{\partial^2 N}{\partial r^2}
+\frac{1}{r^2}\frac{\partial^2 N}{\partial \varphi^2}
+\frac{\partial^2 N}{\partial z^2} 
\right\}
- \Omega(z) v \sigma N.
\label{num:eqnum}
\ee
We start by expanding the desired solution $N$ in a Fourier 
series in $\varphi$ and a Fourier-Bessel series in $r$:
\be
N&=&\frac{1}{\pi} \sum_n \sum_{m}  \left( A_{nm}\cdot\cos \left(n\varphi\right)
+ B_{nm}\cdot\sin \left(n\varphi\right) \right) 
\frac{\jmath_{n}\left( \alpha_{nm} r \right)}
{\left(\jmath_{n}'\left( \alpha_{nm} R \right)\right)^2}
\label{num:ansatznum}
\ee
with $\alpha_{nm}$ being the $m$th solution of $\jmath_{n}\left( \alpha_{nm} R \right)=0$ 
(in ascending order).
We also expand the individual source terms for the point-like sources
\be
S&=& \sum_i q_i(p,t)
\delta\left(\vec{r}-\vec{r}_{i}\right),
\label{num:quelle}
\ee 
where $q_i(p,t)$ is the time and momentum dependence of the particular source and $ \vec{r}_{i}$
its position,  
into the same series 
in $r$ , $\varphi$
\be
S&=&\sum_i q_i(p,t)\delta\left(z-z_{i}\right)
\label{num:ansatznums}
\\
\nonumber
&&\times
\sum_n \sum_{m} \left(\cos \left(n\varphi\right)
 \cos \left(n\varphi_{i}\right)+ \sin \left(n\varphi\right)
 \sin \left(n\varphi_{i}\right) \right)
\frac{\jmath_{n}\left( \alpha_{nm} r \right)\jmath_{n}\left( \alpha_{nm} r_{i} \right)}
{\left(\jmath_{n}'\left( \alpha_{nm} R \right)\right)^2}.
\ee
For extended and continuous sources, e.g. for secondary CRs, one has to use the Fourier-Bessel
representation of the source distribution of the particle in question.

Inserting \Gl{num:ansatznum} and  \Gl{num:ansatznums}
into \Gl{num:eqnum} and using the orthonormality of sine and cosine
 and the analogous property of the Bessel functions  \citep{watson44}

\be
\int_0^R \frac{\jmath_{n}\left( \alpha_{nm} r \right)\jmath_{n}\left( \alpha_{\ n \mu} r \right)}
{\left(\jmath_{n}'\left( \alpha_{nm} R \right)\right)^2} r dr &=& \delta\left(m,\mu\right)
\ee
one obtains  equations for the expansion coefficients $A_{nm}$
\be
\frac{\partial A_{nm} }{\partial t}-S_{nm}^A
 &=& k(p) \left\{ -\alpha_{nm}^2A_{nm}+\frac{\partial^2 A_{nm}}{\partial z^2} \right\} 
- \Omega(z)
v \sigma A_{nm}
\label{num:eqnuma}
\ee
with 
\be
S_{nm}^A&=&\sum_i q_i(p,t)  \cos \left( n \,\varphi_{i} \right)
\frac{\jmath_{n}\left( \alpha_{nm} r_{i} \right)}
{\left(\jmath_{n}'\left( \alpha_{nm} a \right)\right)^2}
\ee
and  similar equations for $B_{nm}$ 
\be
\frac{\partial B_{nm} }{\partial t}-S_{nm}^B
 &=& k(p) \left\{ -\alpha_{nm}^2B_{nm}+\frac{\partial^2 B_{nm}}{\partial z^2} \right\} 
- \Omega(z)
v \sigma B_{nm}
\label{num:eqnumbb}
\ee
with 
\be
S_{nm}^B&=&\sum_i q_i(p,t)  \sin \left( n\, \varphi_{i} \right)
\frac{\jmath_{n}\left( \alpha_{nm} r_{i} \right)}
{\left(\jmath_{n}'\left( \alpha_{nm} a \right)\right)^2}.
\ee
These equations are not analytically solvable
for arbitrary  $\Omega(z)$, therefore one has to use numerical methods. 
The advantage of this ansatz is, that it is possible to solve these one-dimensional
PDEs simultaneously on many computers.
Also, the resolution at a given point obtained in $r, \varphi$ merely  depends on the number
of coefficients in the series ansatz, \Gl{num:ansatznum}, that is actually computed.

As the $\alpha_{nm}$'s increase monotonically with  $n$ and also with $m$,
one sees from  \Gl{num:eqnuma}, that for large $m,n$ we have 
$
 k(p)\alpha^2 \gg \Omega(z) v \sigma B_{nm}
$
and therefore, the latter term may be neglected. In this case an analytical solution is known. 

\subsection{CR distribution due to a single source}
Before we start solving \Gl{num:eqnuma} and \Gl{num:eqnumbb}, we have to determine the number
of coefficients to be  taken into account in the ansatz \Gl{num:ansatznum}.
 Evidently, the spatial resolution in $r,\, \varphi$ direction of
the solution obtained by the ansatz    \Gl{num:ansatznum} depends on the 
number of coefficients used and also on the distance $r$ from the origin.
As we are mainly interested in the CR density in the vicinity of the Sun,
we determine the number of coefficients by comparing the solutions 
of the propagation equation \Gl{num:eqnum} obtained by a truncated series according to ansatz  
\Gl{num:ansatznum} with
the solution of the propagation equation for a spherical source with a radius of $\rho_s\,=\,50$\,pc, located at
the position of the Sun.
{To ease the calculations we neglect the geometry of the Galaxy, assuming 
the loss processes not to depend on the spatial coordinates. This is 
permitted if we consider only the vicinity of the source 
for a limited time after the SN explosion.
So, we study the source in an uniformly distributed ISM.} 

First, we derive a solution  for a spherical source.
In this case, placing the source at the origin of our coordinate system 
provides us with spherical symmetry,
i.e. the  solution only
depends on the radial coordinate, $\rho$.
We further restrict ourselves to one fixed particle momentum $p$.
Then \Gl{gle:mod} can be written as
\be
\frac{\partial N}{\partial t}
-S
=
k\left(\frac{\partial^2 N}{\partial \rho^2}+
\frac{2}{\rho}\frac{\partial N}{\partial \rho} \right)
-b N
\label{disc:kugelks}
\ee 
with $b=\Omega \,v\,\sigma$. {}
The source $S$ has the temporal form given in \Gl{rech:sstrength}, 
so we have
\be
S&=&\hat q_i\cdot \left(t-t_i\right){\rm exp}\left(-\frac{t-t_i}{\tau}\right) 
\Theta\left(t-t_i\right) 
\Theta\left(\rho_s-\rho\right).
\label{disc:source}
\ee
Without loss of generality, we set $\hat q_i\,=\,1$ and $t_i\,=\,0$. 

For \Gl{disc:kugelks} one can find the Green's function
\be
G&=& \frac{1}{8\pi\sqrt{\pi}\sqrt{k}}\frac{1}{\rho\,\rho_0\sqrt{t-t_0}}
{\rm exp}\left(-b\left(t-t_0\right)\right)
\nonumber \\ 
&&\times{\rm exp}\left(-\frac{\rho^2+\rho_0^2}{4\,k\left(t-t_0\right)}\right)
{\rm sinh}\left(\frac{2\,\rho\,\,\rho_0}{4k\left(t-t_0\right)}\right).
\label{dics:greenk}
\ee
For a verification that \Gl{dics:greenk} is indeed a Green's function
for \Gl{disc:kugelks} and a short discussion 
on how to obtain it, we refer to Appendix \ref{app:appc}.

\begin{figure}
\includegraphics[width=\textwidth]{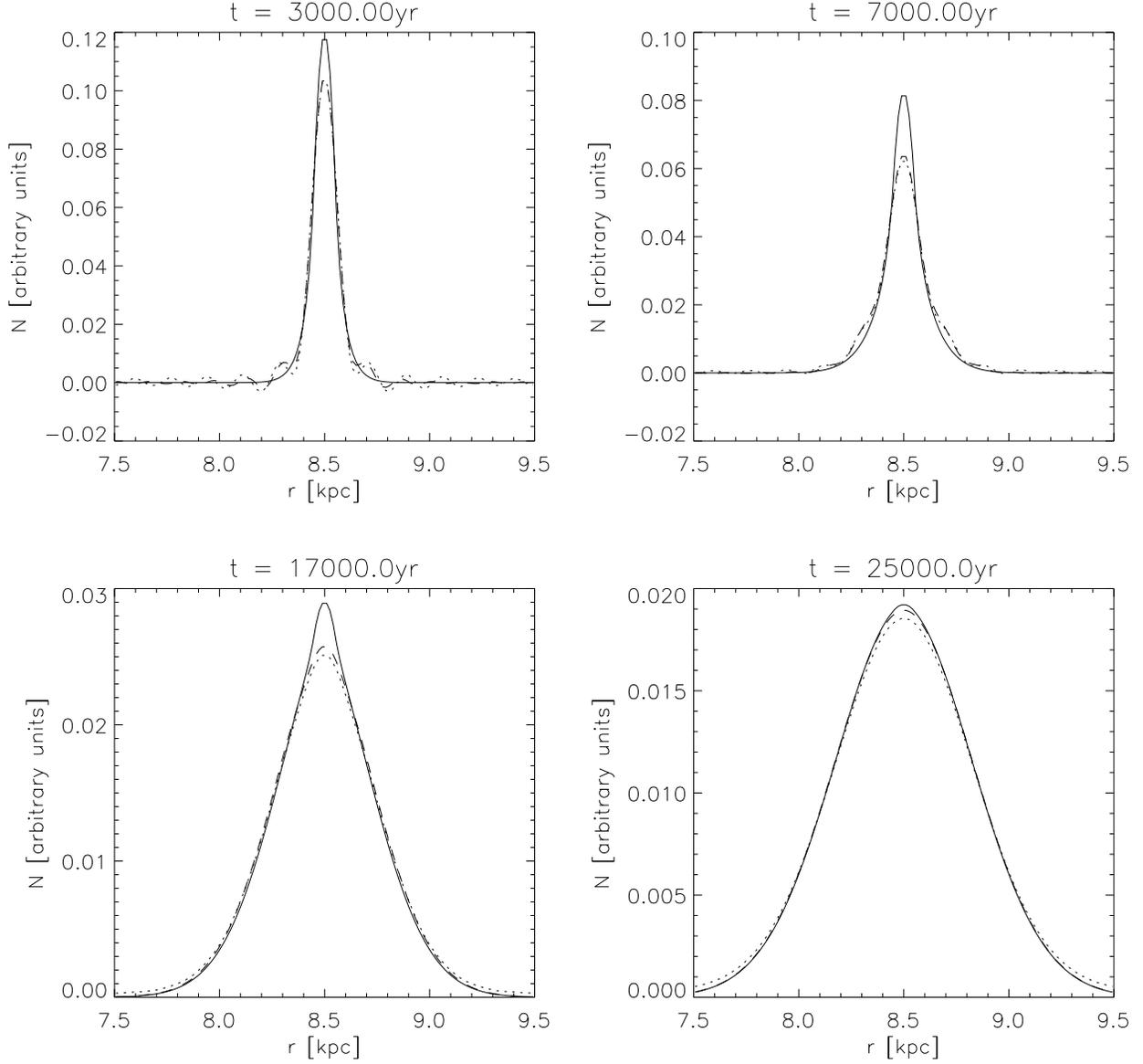}
\caption{Comparison of CR density due to a spherical source with radius 50\,pc
given by \Gl{numi:vglint} (solid line) and that computed using \Gl{num:ansatznum} with 312 coefficients
used for the series in $n$ and 210 coefficients used in the series in $m$;
 cuts in  $r$ (dotted line) 
and $\varphi$ (dashed line) direction, 
for different times  from source ``ignition''. 
}
\label{fig:qvergl}
\end{figure}

Thus, we obtain the solution of \Gl{disc:kugelks} with sources \Gl{disc:source}
by the convolution 
\be
N\left(\rho,t\right)&=&\int_{\rho_0=0}^{\rho_s}\int_{t_0=0}^t G\left(\rho,\rho_0,t,t_0\right)\cdot 
S\left(\rho_0,t_0\right)\,d\rho_0dt_0\\
&=&\int_{\rho_0=0}^{\rho_s}\int_{t_0=0}^t \frac{1}{8\pi\sqrt{\pi}\sqrt{k}}\frac{1}{\rho\,\rho_0\sqrt{t-t_0}}
{\rm exp}\left(-b\left(t-t_0\right)\right)
{\rm exp}\left(-\frac{\rho^2+\rho_0^2}{4\,k\left(t-t_0\right)}\right)
\nonumber \\
&&\times 
{\rm sinh}\left(\frac{\rho\,\,\rho_0}{2\,k\left(t-t_0\right)}\right)
  \cdot \,t_0{\rm exp}\left(-\frac{t_0}{\tau}\right) \nonumber \\ 
&&\times \Theta\left(t_0\right) 
\Theta\left(\rho_s-\rho\right)
\rho_0^2d\rho_0 dt_0.
\label{diss:solcon}
\ee
The  $\rho_0$ integration can be performed analytically, which leads to 
\be
N\left(\rho,t\right)&=&\int_{t_0=0}^t 
\left\{ -2\,\sqrt {k \left( t-{ t_0} \right) }{\,\,\exp\left({{\frac {
-{\rho_s}^{2}-{\rho}^{2}}{4\,k \,\left( t-{  t_0} \right) }}}\right) }\,\sinh \left( {
\frac {\rho_s \rho}{2\,k \left( t-{  t_0} \right) }} \right) {\rho}^{-1} \right.
 \nonumber
\\
&&
\left.
+\frac{\sqrt {\pi }}{2} \left( {\rm  erf} \left( \frac{1}{2}\,{\frac {\rho_s-\rho}{\sqrt {k \left( t-{  t_0}
 \right) }}} \right) +{\rm  erf} \left( \frac{1}{2}\,{\frac {\rho_s+\rho}{\sqrt {k
 \left( t-{  t_0} \right) }}} \right)  \right)  \right\} 
\nonumber \\
&& \times
{\exp\left({b \left( 
{  t_0}-t \right) }\right)}{ t_0}\,{\exp\left(-\frac{ t_0}{\tau}\right)}dt_0 .
\label{numi:vglint}
\ee
Unfortunately, this integral  is not solvable analytically.
It was computed  numerically using a Riemann sum.
The result is plotted in \Fig{fig:qvergl}, where we also plotted
the solution of \Gl{gle:mod}, using  addends with $m\leq210$ and $n\leq311$ in ansatz
\Gl{num:ansatznum}, which turned out to be the best
tradeoff of numerical complexity versus
spatial accuracy. 
{This solution using ansatz \Gl{num:ansatznum} describes well, both in $r-$ and $\varphi$-direction,
the spatial evolution of the
CR density around a spherical source.}

\section{Signatures  of Discrete Sources in the Galactic Cosmic Ray Distribution}
\label{sec:signatures1}
In this section we  represent first results 
of our investigation as to what extent the SN origin of CR affects the density 
distribution of Galactic CR.
We  study the temporal evolution
and the fluctuations in the CR density,
using the methods developed in the previous section.

\subsection{Randomly Distributed Sources}
\label{sec:usdi}
We now consider  the time-dependent propagation equation \Gl{num:eqnum} 
that  we  solve using the ansatz  \Gl{num:ansatznum} for several source distributions. 
 As a first step we performed a calculation for CR point sources that are 
randomly distributed in the Galactic plane. 
For that purpose we chose $^{16}$O, for it is regarded as the most abundant primary
CR element beyond helium.  
For this calculation and those described in the following sections,
the CR density $N$ was computed using  ansatz \Gl{num:ansatznum}.
The corresponding coefficients $A_{nm}$, $B_{nm}$
were obtained numerically using a semi-implicit scheme based on the  
Du Fort-Frankel \citep{dufort53} and leapfrog schemes.
We started 
the calculation of the temporal evolution of the CR density ,$N$, 
from the appropriate steady state solutions, 
 which were calculated using the  package  TOMS638 \citep{houstis85a,houstis85b}.

For the source function we use $n_q$ randomly distributed point sources with a
radial probability  distribution given by  \Gl{rech:sourcedis}. Considering only one 
CR primary element at a time, we have for the 
source term $S$ in \Gl{num:eqnum}: 
\begin{eqnarray}
S(r,\varphi,z,t)&=&\sum_{i=1}^{n_q} q_i \left( t,p \right) \delta \left({\vec{r}}-{\vec{r}_i}\right)
\end{eqnarray}
with 
the source strengths $q_i$ defined in \Gl{rech:sstrength},
and
$\vec{r}_i$  the  locations of the $i$th source.
The number of sources, $n_q\,=\,150000$, was chosen to reproduce the local supernova surface density of 
25 SN per Myr and kpc$^2$ of the Galactic disk \citep{grenier00}.
We calculated CR densities for several  random SN distributions in the Galactic disk, each for a time interval   
of $10\,$Myr.

\begin{figure}[pH!]
\begin{tabular}{l}
\begin{minipage}{\columnwidth}
\includegraphics[width=0.7\columnwidth]{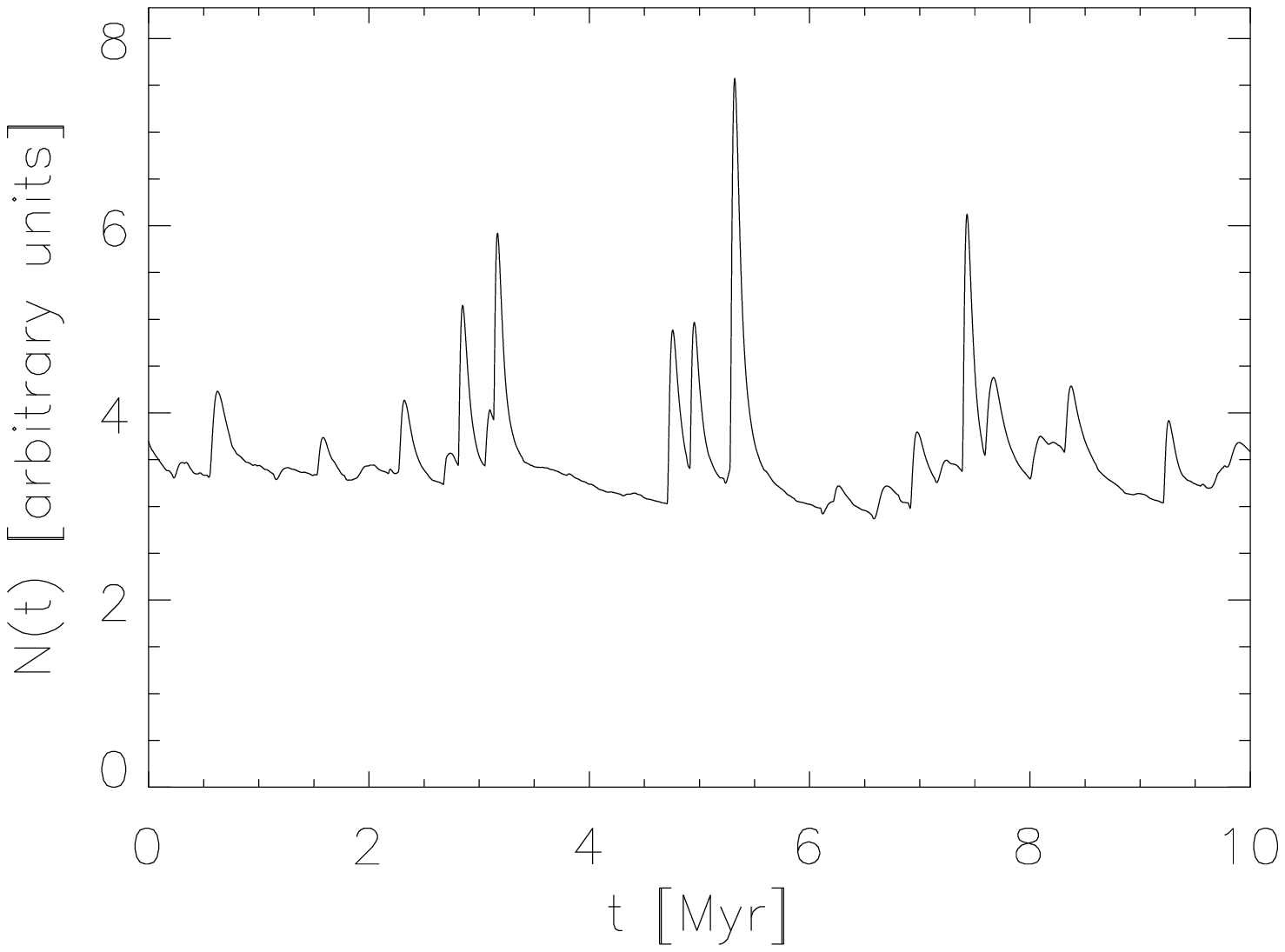}
 \end{minipage}
\\
      \begin{minipage}{\columnwidth}
\includegraphics[width=0.7\columnwidth]{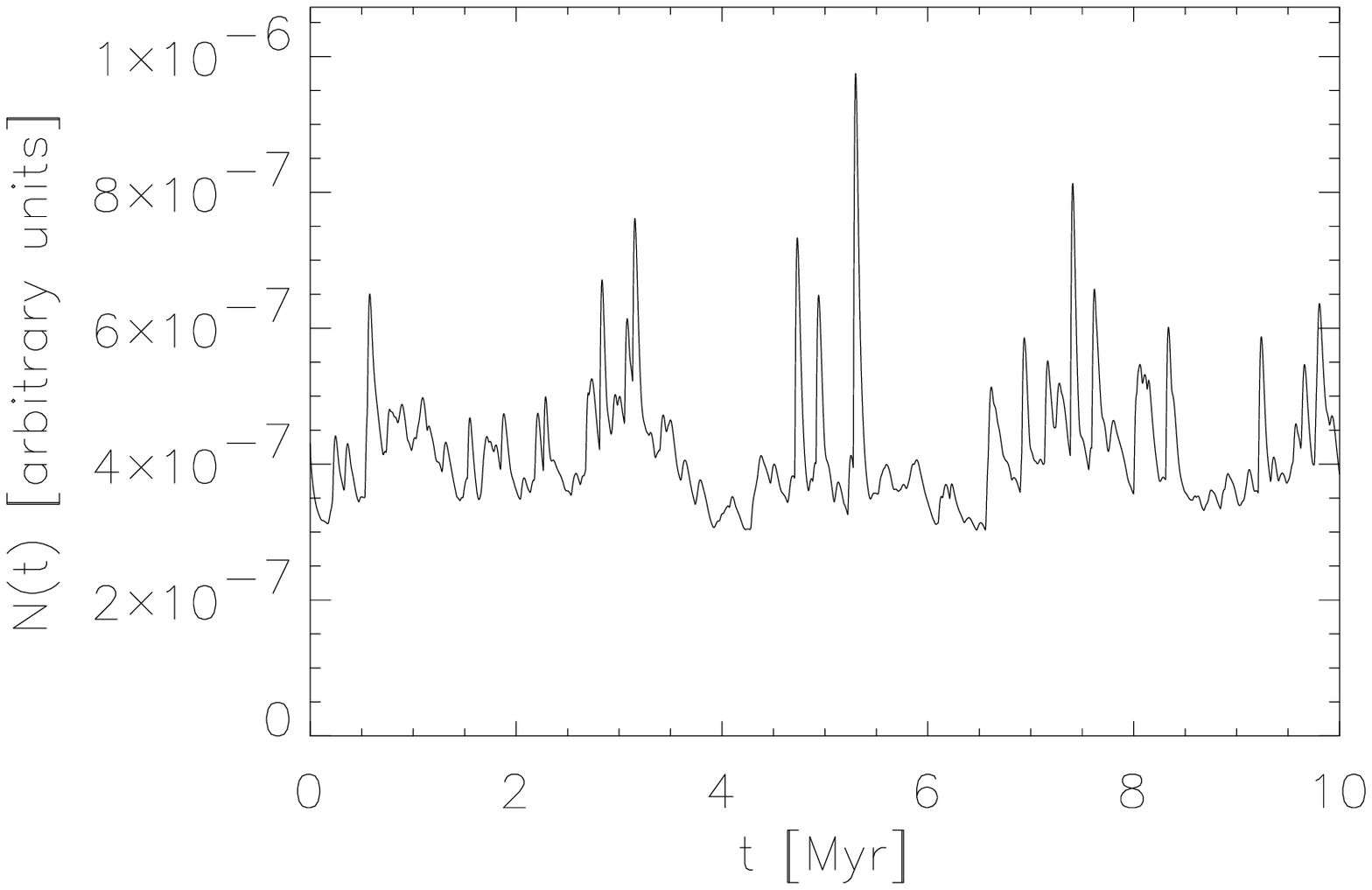}
\vspace{.0pc}
      \end{minipage}
\\
\end{tabular}
\caption{Temporal variation of the $^{16}$O CR primary density at the position of the Sun, for
10\,GeV per nucleon  (upper panel) and 5\,TeV per nucleon (lower panel).
}
\label{fig:time16o}
\end{figure}

\begin{figure}[pH!]
\vspace{-1cm}
\includegraphics[width=0.7\columnwidth]{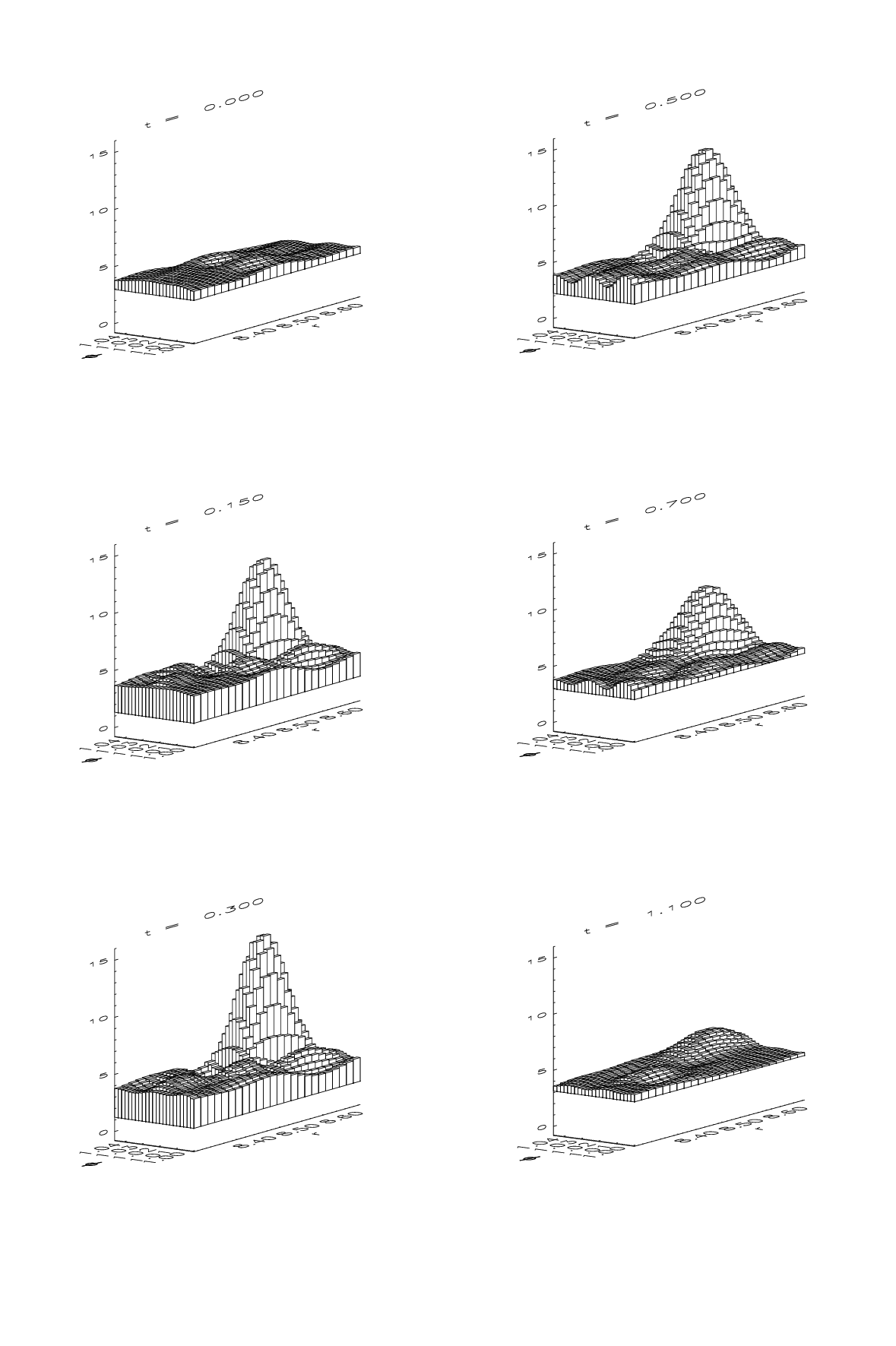}
\\
\vspace{-4cm}
\\
\caption{
The density of $^{16}$O at $E=10\,$GeV per nucleon in  a 400\,pc $\times$
400\,pc section  
of the Galactic plane ($z=0$) 
during a local SN event  
for several times (in Myr). 
The Sun ($r=8.5\,$kpc, $\phi=1.025$) is positioned at the center. 
The local
CR density is strongly influenced by nearby SN, although the excess quickly
disappears.
}
\label{fig:figure_dichte6}
\end{figure}

In these calculations, the coefficients with  
$1\leq m \leq 210$
and
$0\leq n \leq 311$
of the series 
ansatz \Gl{num:ansatznums} were computed. This
yields  a resolution in $r$, $\varphi$ of  $\sim$ 170\,pc at the position of the Sun. 
The grid spacing in $z$     
 is 20\,pc, the time step is 1\,kyr.

Computing the series \Gl{num:ansatznums} at the position of the Sun gives  the time variation 
of the CR density at this position, as is shown in \Fig{fig:time16o},
where  the CR density for an energy of 10\,GeV per nucleon (upper panel)
and for an energy of 5\,TeV per nucleon (lower panel) is plotted versus time. 
These  figures illustrate
 that the density of  CR with an energy of 5\,TeV 
 shows more rapid fluctuations
than at an energy of  10\,GeV  on account of the energy
dependence of the diffusion coefficient.

The variations in the CR density at a given location, shown in \Fig{fig:time16o}, 
have a typical amplitude of about
20 per cent with occasional spikes reaching 100 per cent.
The latter mainly occur due to nearby SN explosions, as
illustrated in \Fig{fig:figure_dichte6}, where the temporal evolution of the CR 
density  at an energy of 10\,GeV per nucleon is shown for 
a 400\,pc$\, \times  \,$400\,pc section of the Galactic plane.
  Variations of the same order of magnitude of the local CR energy density  due to nearby SN have been
  found by \citet{lingenfelter69} when he investigated the contribution of 
these SN to the CR energy density near our Sun, using a simple diffusion type 
model only considering losses due to escape.
The perturbation in the CR density due to the source 
stays almost localized and does not spread out.

\subsection{The influence of Gould's Belt Star-Burst Region}
\label{sec:gbrechnen}

The local distribution of stars and the ISM 
is dominated by 
 the  so-called Gould's Belt  \citep{poeppel97}.
Of particular interest for our investigations is the enhanced SN rate within
the belt \citep{grenier00}.
The kinematics of the belt can be modeled by an expanding ring of gas,
 assuming an initial explosive event
\citep{perrot03}.
Its age is estimated to be about 30\,Myr which implies that
with regard to our calculations, which cover the last 10\,Myr, not only its presence, 
but also the still evolving 
geometry of the belt has to be taken into account.
 We assume that the enhanced supernova rate applies to the entire volume of Gould's belt, i.e.
we neglect a possible delay on account of the main-sequence phase of very massive stars. The probability distribution
for supernova explosion would then be given by the time-dependent expansion model of
\citet{perrot03} in addition to the stationary large-scale SN distribution in the Galaxy
(see Eq.\ref{rech:sourcedis}).

\begin{figure}[p]
\begin{tabular}{l}
\begin{minipage}{\columnwidth}
\includegraphics[width=0.7\columnwidth]{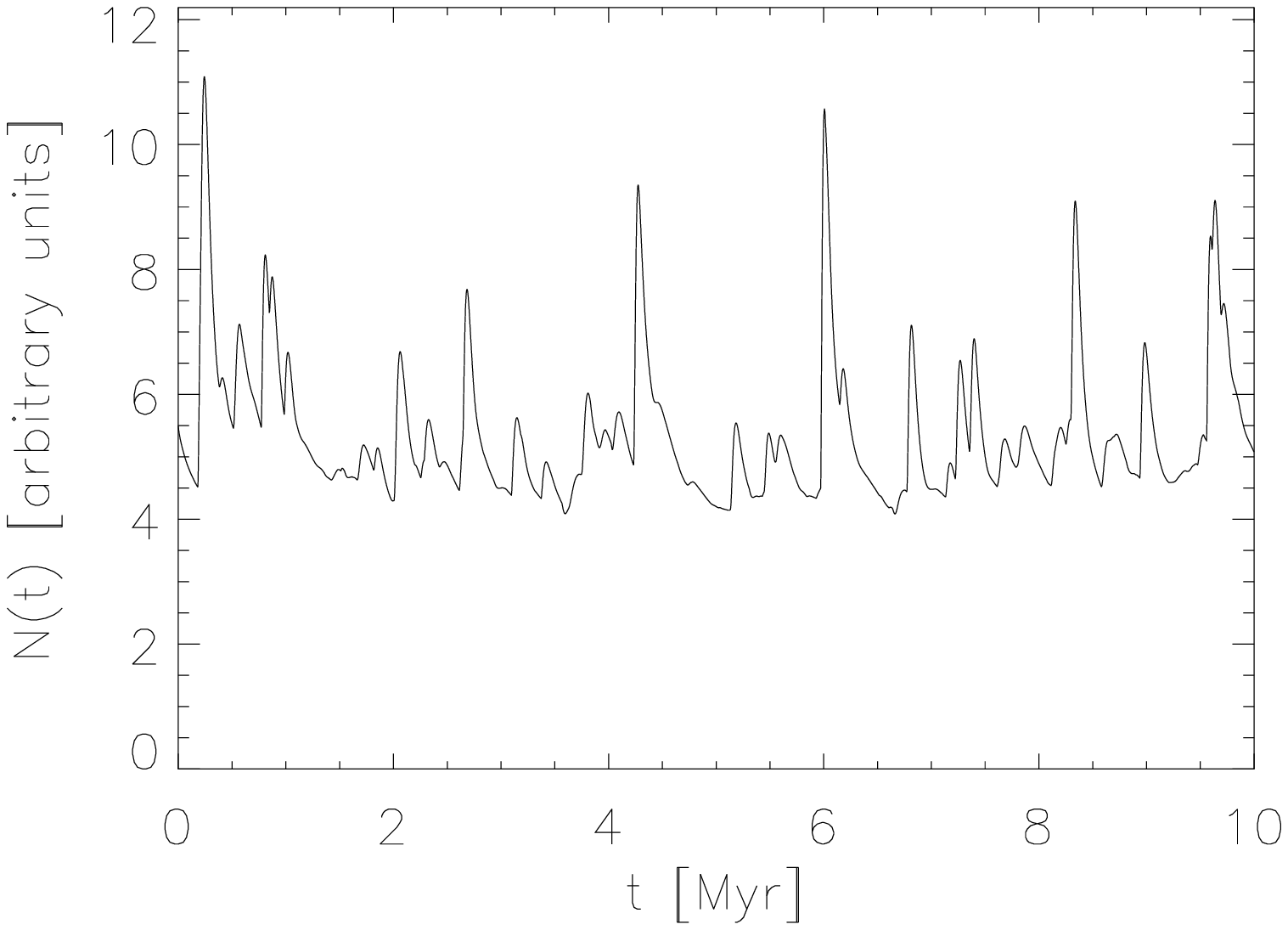}
 \end{minipage}
%      \hspace{1pc}
\\
      \begin{minipage}{\columnwidth}
\includegraphics[width=0.7\columnwidth]{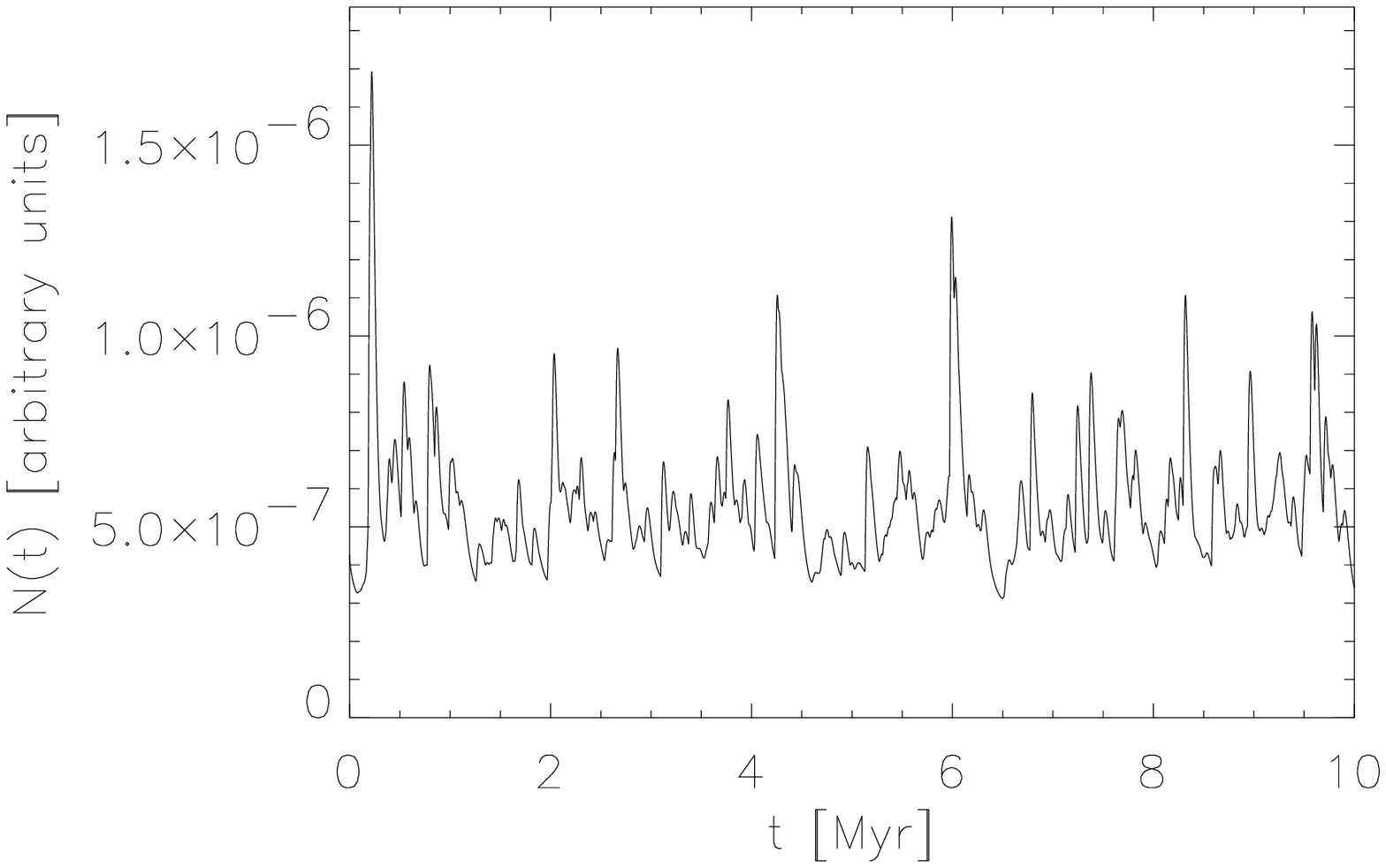}
\vspace{.0pc}
      \end{minipage}
\\
\end{tabular}
\caption{ Same as  \protect \Fig{fig:time16o},  assuming a locally enhanced SN rate  
in  Gould's Belt. 
 }
\label{fig:tschnittmgb}
\end{figure}

The results  displayed in \Fig{fig:tschnittmgb} 
show an enhanced variation of the local CR density due to the
locally increased SN rate compared to the case without Gould's Belt,
 that has been given in the last section and is shown in \Fig{fig:time16o}.  
The locally enhanced SN rate  also  leads to a locally
increased 
mean CR density which is most pronounced at low energies  or,
in other words, SNR need less efficiency as CR accelerators to 
replenish the galactic cosmic rays. 

\subsection{Secondary Cosmic Ray Particles}
In  section \ref{sec:usdi} we showed that the density of CR primary nuclei 
 may vary up to about
one order of magnitude in space and time. 
To investigate to 
what extent this is also true for secondary CR nuclei and, therefore,
whether there are any  effects on
the  ratio of secondary to primary CR isotopes, we further performed 
 calculations including secondary nuclei.
As the ratio of Boron to Carbon is widely used to quantify the parameters
of various propagation models, we considered 
 the isotopes $^{12}$C, which was assumed to be pure primary and $^{11}$B
which was assumed to be produced solely by interactions of the primary $^{12}$C
with the interstellar matter. The results of these calculations are shown 
\begin{figure}[p]
\begin{tabular}{l}
\begin{minipage}{\columnwidth}
\includegraphics[width=0.7\columnwidth]{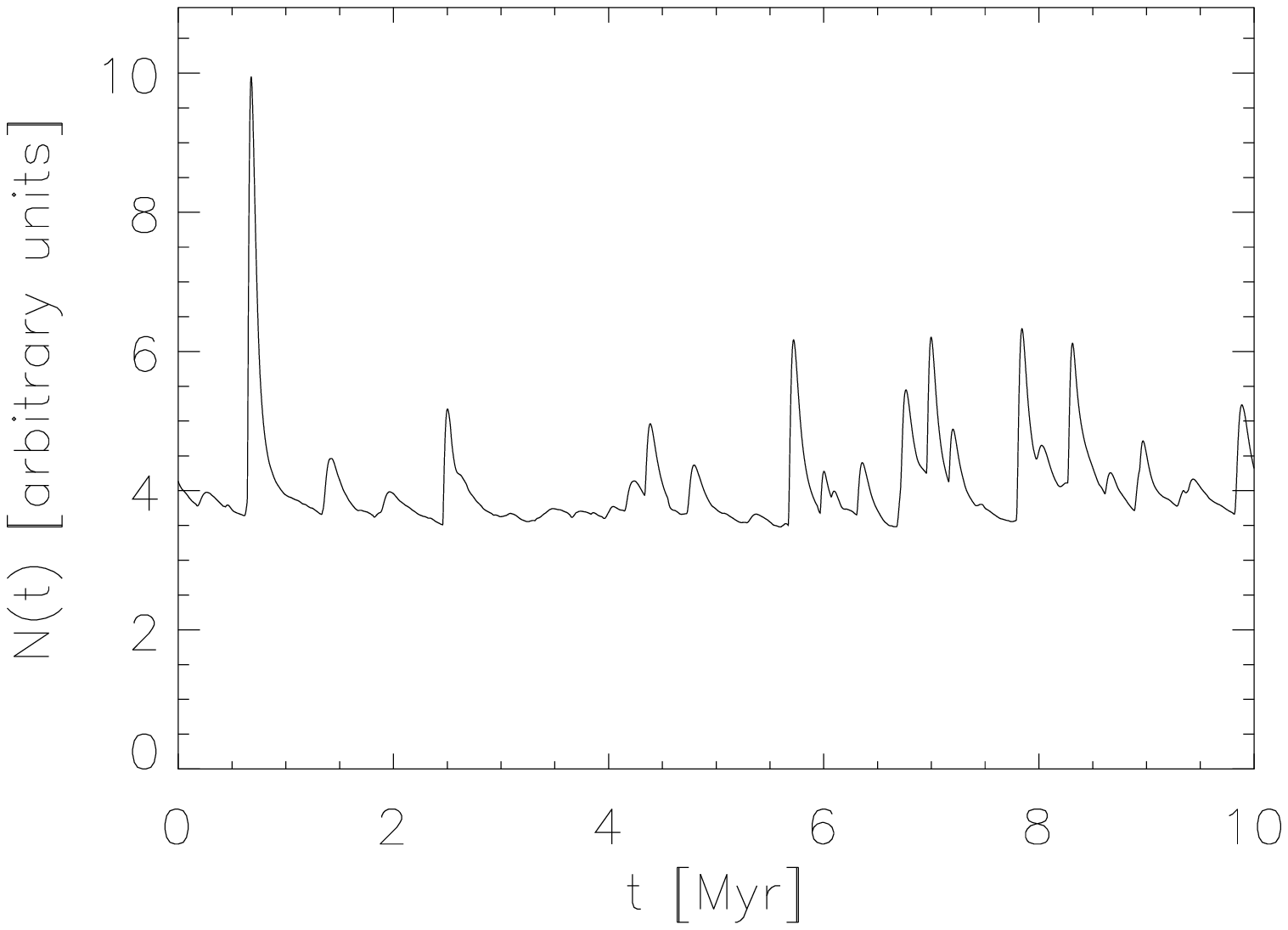}
 \end{minipage}
\\
      \begin{minipage}{\columnwidth}
\includegraphics[width=0.7\columnwidth]{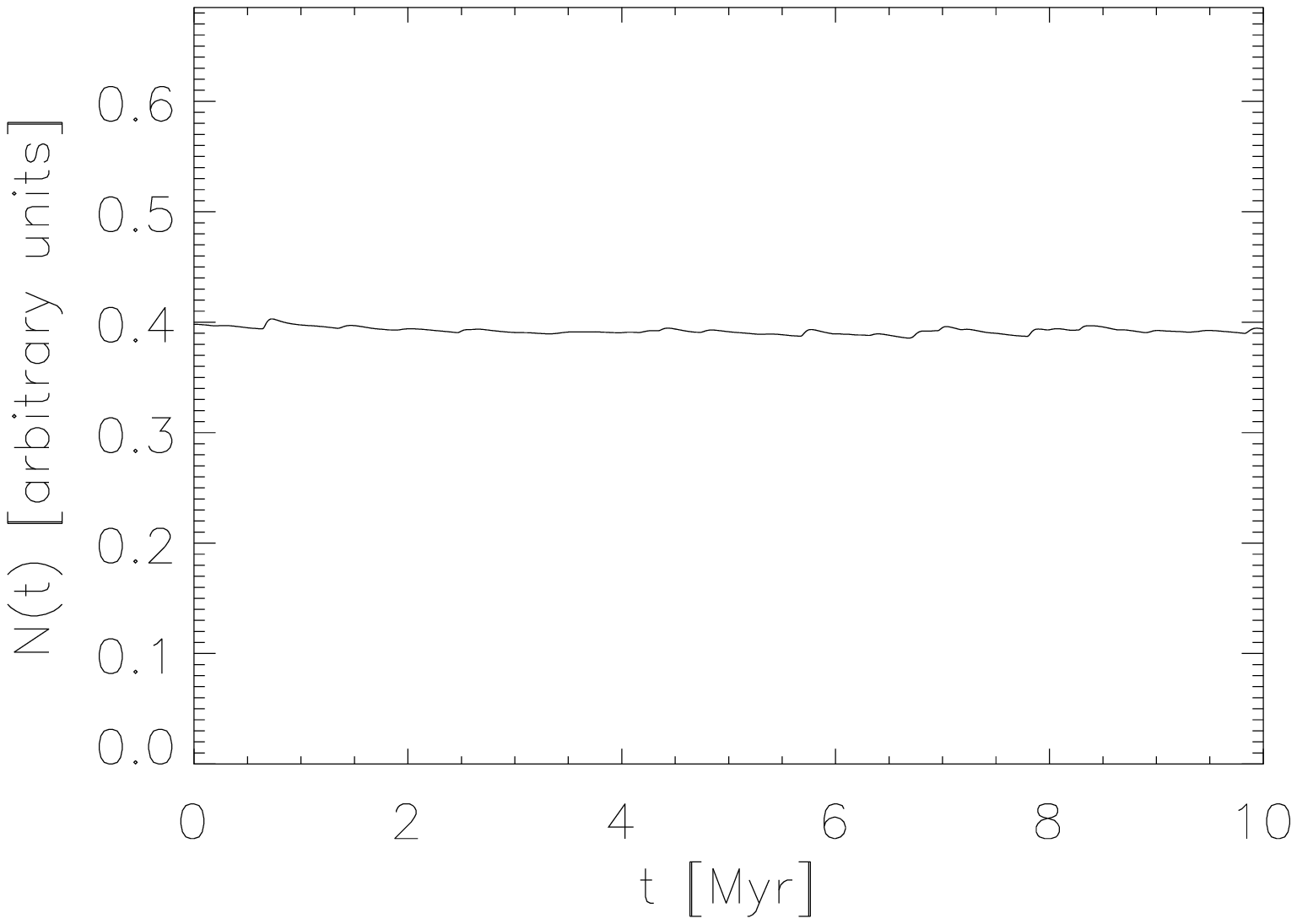}
\vspace{-1.0pc}
      \end{minipage}
\\
\end{tabular}
\caption{Temporal variation of the CR primary density $^{12}$C (upper panel) and secondary $^{11}$B
(lower panel) at the position of the Sun at energy of 10\,GeV per nucleon.
The deviation from the 
average value  for the CR secondary density  is way smaller
than that for the  CR primary density.
}
\label{fig:time12c}
\end{figure}
in \Fig{fig:time12c}, where we plot, as in \Fig{fig:time16o}, the CR density 
at 10\,GeV per nucleon versus time 
for the primary isotope $^{12}$C (upper panel) and the secondary isotope
  $^{11}$B (lower panel), both at 
 the position of
the Sun. 
\Fig{fig:time12c} shows
that, although the density of the primary CR varies  by a factor of 2, as expected 
from the previous calculations,  
the variation of the secondary CR particles
is only a few per cent. This statement holds true also for higher particle energies.
\begin{figure}
\begin{tabular}{c c}
\begin{minipage}{.45\columnwidth}
\includegraphics[width=1.1\textwidth]{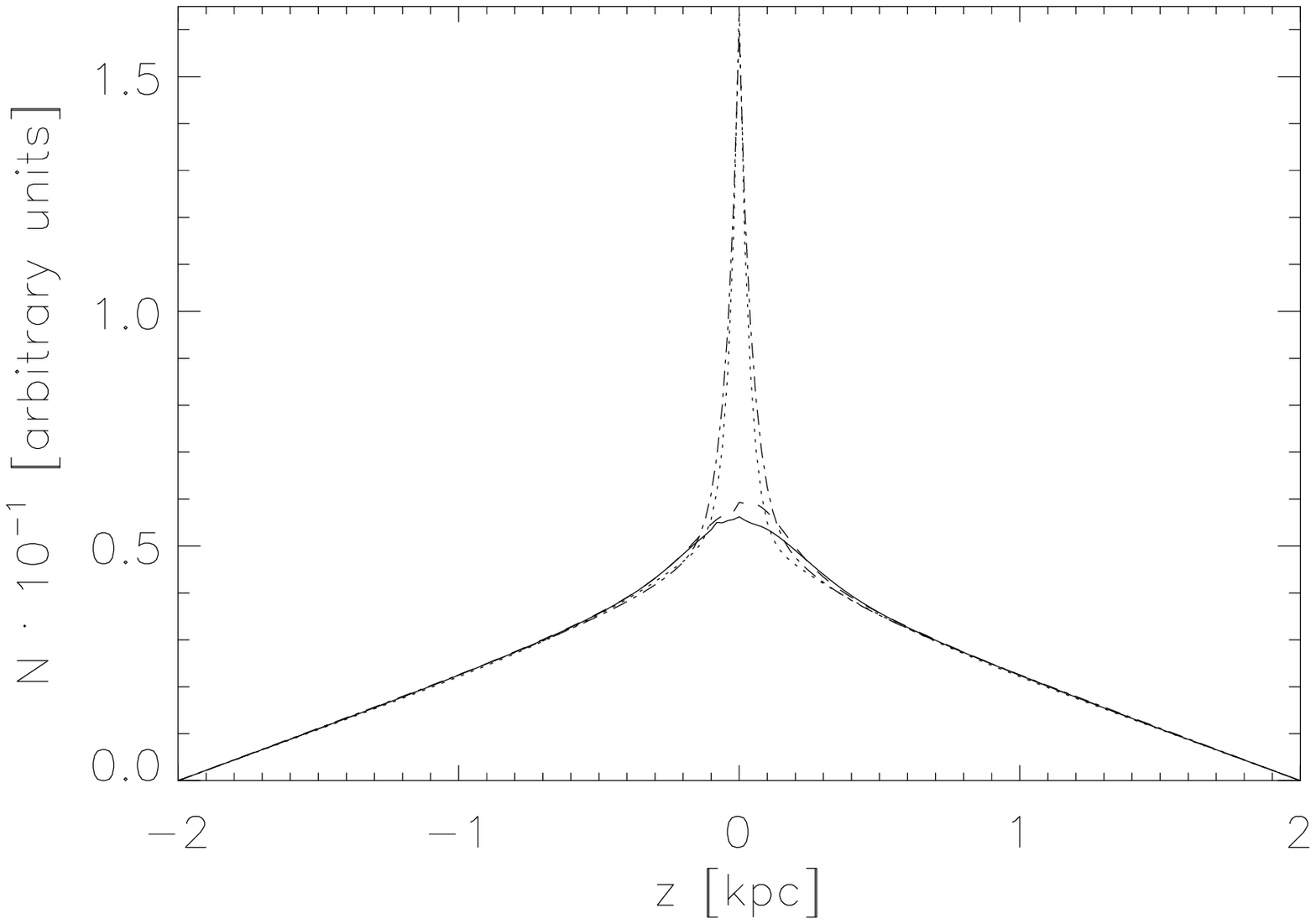}
\includegraphics[width=1.1\textwidth]{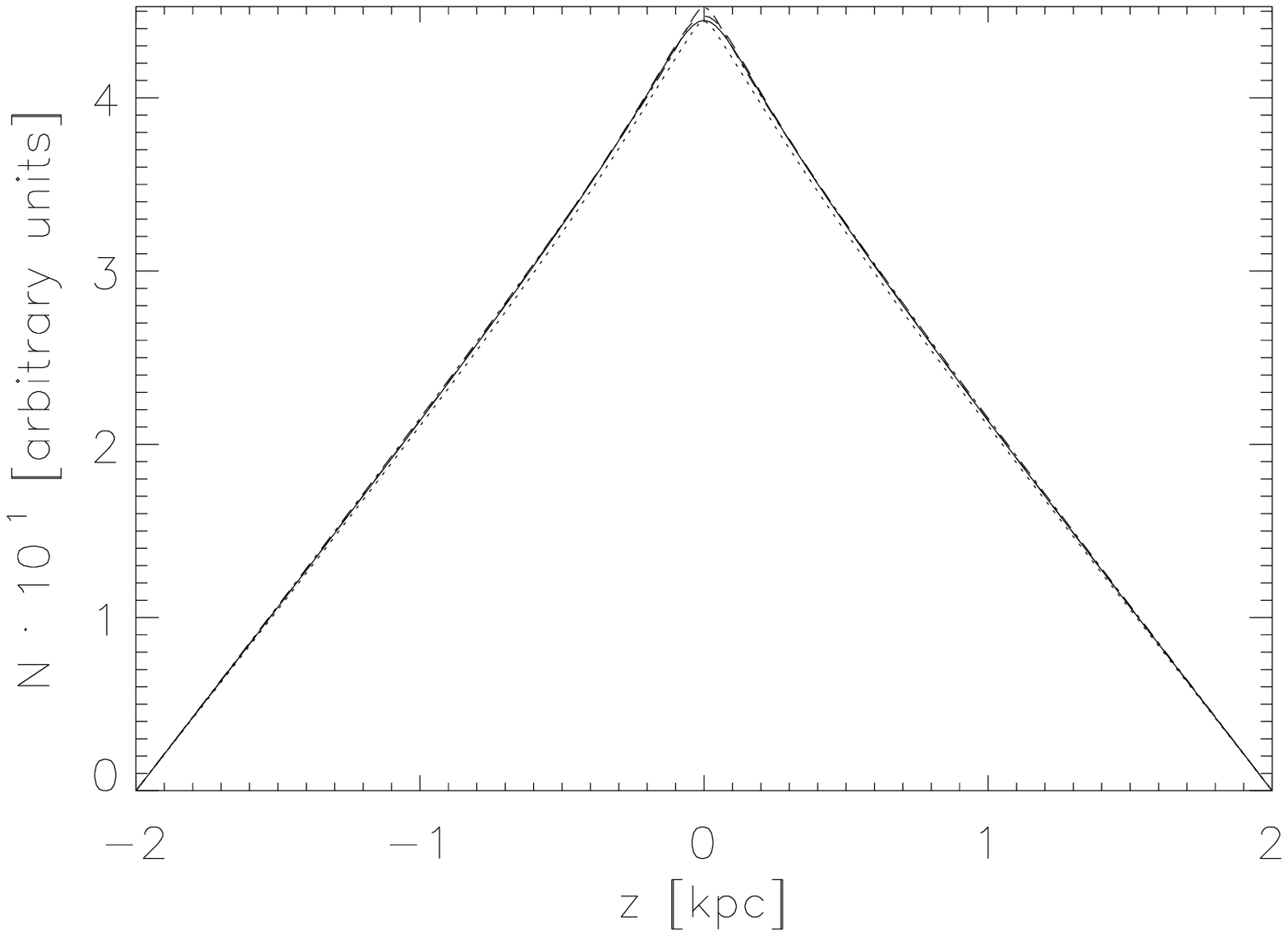}
\end{minipage}
&
\begin{minipage}{.45\columnwidth}
\includegraphics[width=1.1\textwidth]{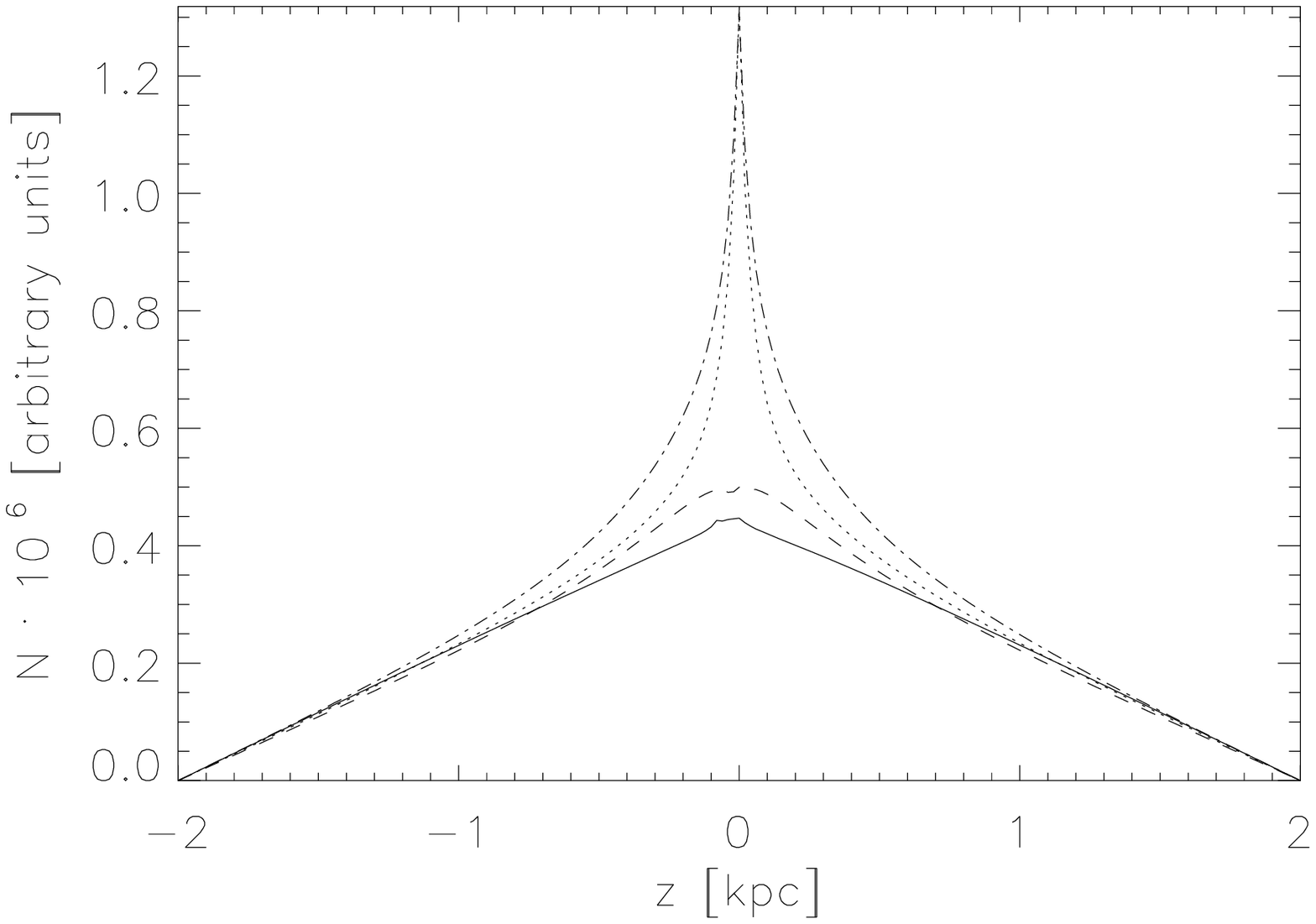}
\includegraphics[width=1.1\textwidth]{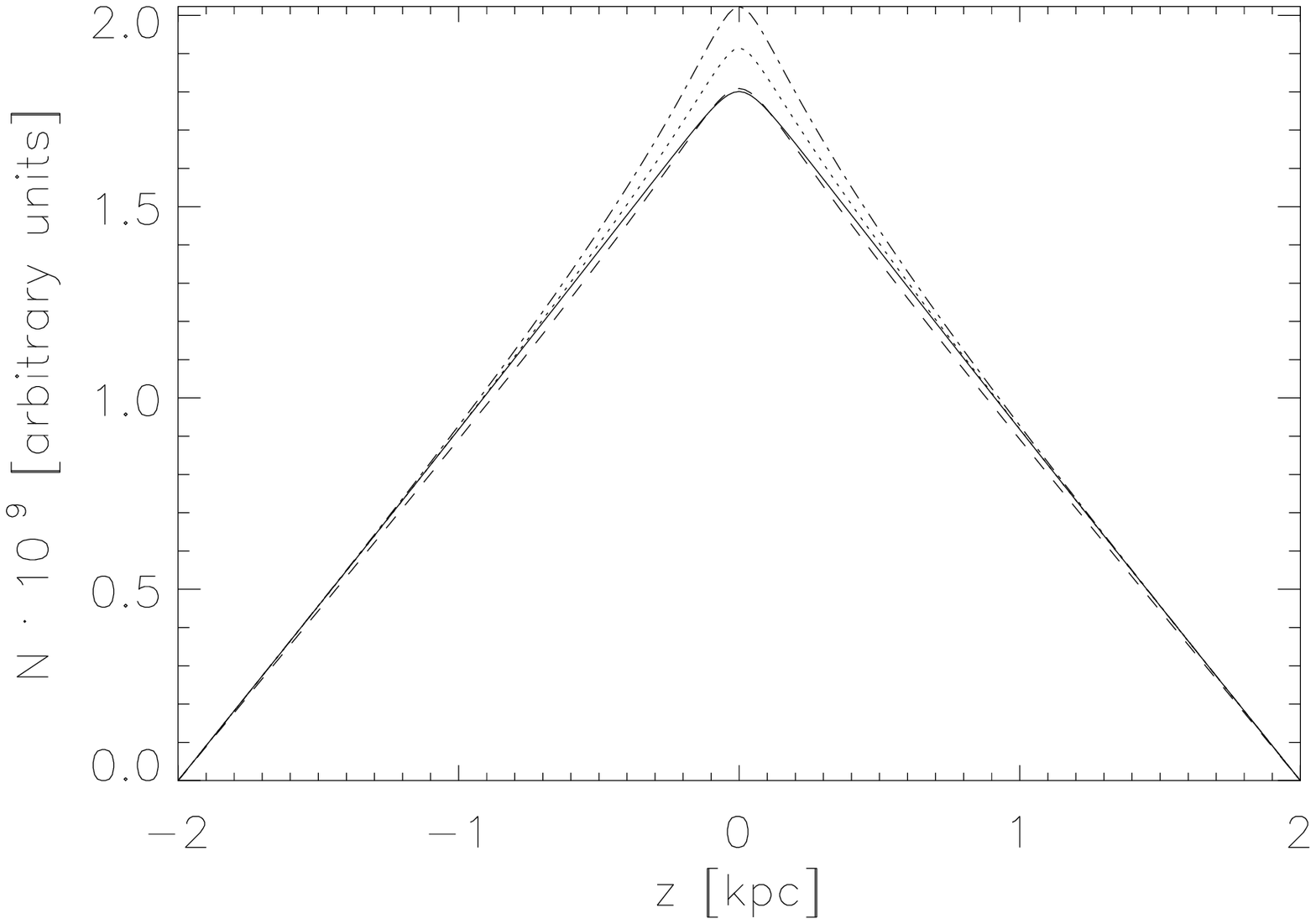}
\end{minipage}
\\
\end{tabular}
\caption{ The density of primary
CRs (upper panel) and that of secondary CRs
(lower panel) plotted as a function of distance from the Galactic plane
at $R_0=8.5$\,kpc from the Galactic center. The left column is for a particle
energy of 10\,GeV per nucleon, and the right column applies to particles at 5\,TeV energy.
We show the density distributions at four {arbitrarily chosen}  instances of time, indicated by different
 line-styles.
}
\label{fig:zschnittebc}
\end{figure}

Having analyzed the variations of the CR density with time (\Fig{fig:time16o},
  \Fig{fig:time12c}) and in the Galactic plane (\Fig{fig:figure_dichte6}),
we now are interested in the variations of the CR flux perpendicular to the
Galactic disk. 
In \Fig{fig:zschnittebc},
we compare the densities of primary (upper panels) and secondary (lower panels) CR
 perpendicular to the Galactic plane, for CR energies
of 10\,GeV (left panels) and 5\,TeV (right panels), respectively, 
 at four different instances of time, indicated by different line-styles. 

 As in the case of \Fig{fig:time12c}, where we
plotted the  density of CR primary and secondary particles at one position in the Galactic plane versus time,
these figures show, that there is only a marginal variation of the CR secondary 
distribution perpendicular to the Galactic plane, despite rather large variations 
in the CR primary distributions. This behavior depends only
weakly on the particle energy. Nevertheless, for particle energies of 
5\,TeV, there is a variation in the  density of CR secondaries at $z=0$ of roughly 10\%.  
These findings have far-reaching consequences for the ratio of secondary to primary CR, as 
discussed in the next section.

\section{Global signatures of an SNR origin of cosmic rays}
\label{sec:disskuss}

 Equipped with a technique to calculate
the spatial and temporal distribution of CR primary and secondary
elements with SN-like objects as sources, we  now 
 discuss the observational implications for CR astrophysics.
We discuss the calculated spectra themselves and also compare them 
with measurements for the  CR primary elements 
$^{16}$O and $^{12}$C.

To obtain a measure of the variations of the CR spectra  at the position of the Sun, 
we simulated 240 possible CR spectra both for the standard Galactic plane distribution of 
supernovae as given by Eq.\ref{rech:sourcedis} and for the Gould's belt supernovae
in addition to that.

\begin{figure}
\begin{tabular}{l}
\begin{minipage}{\columnwidth}
\includegraphics[width=0.7\columnwidth]{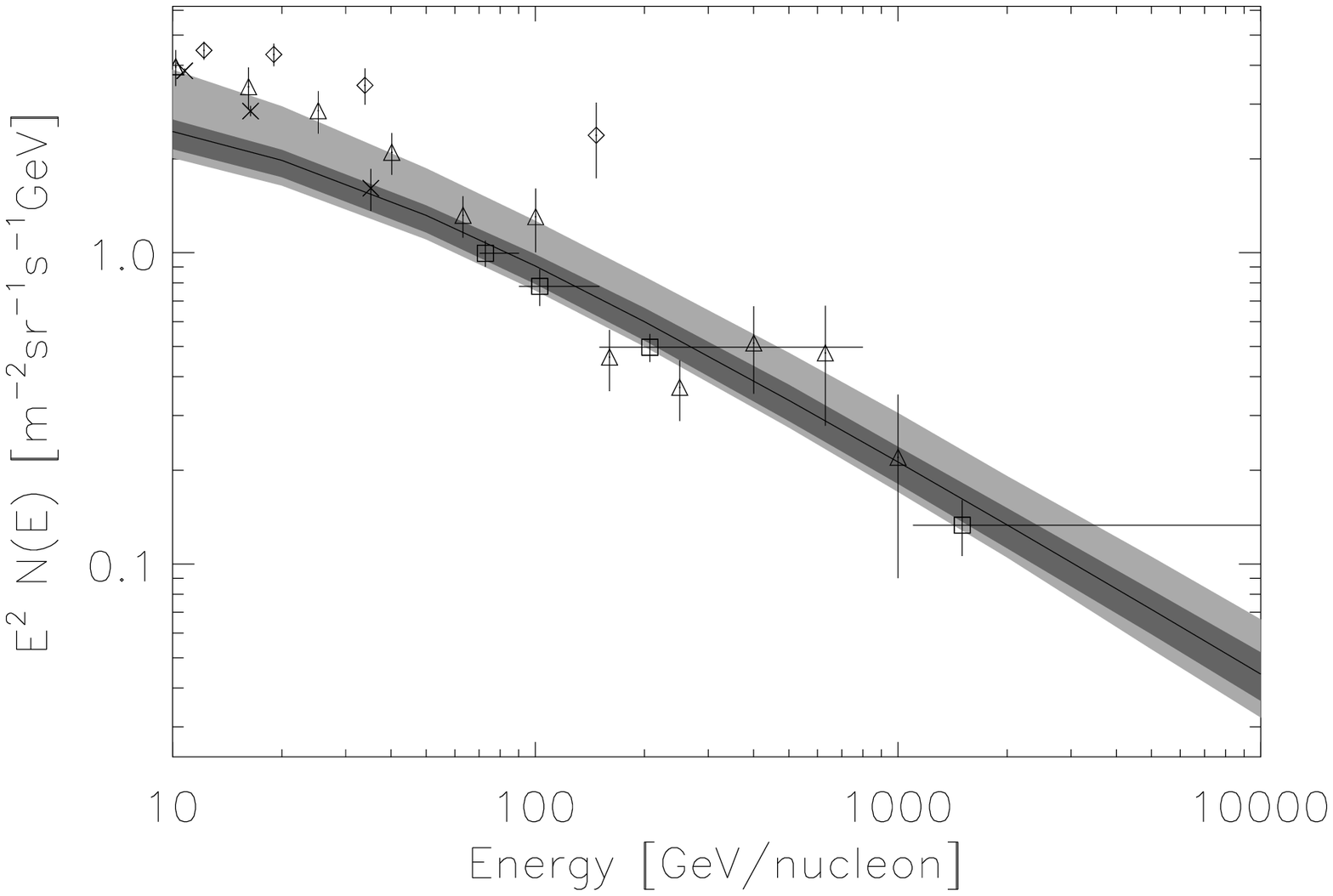}
 \end{minipage}
\\
      \begin{minipage}{\columnwidth}
\includegraphics[width=0.7\columnwidth]{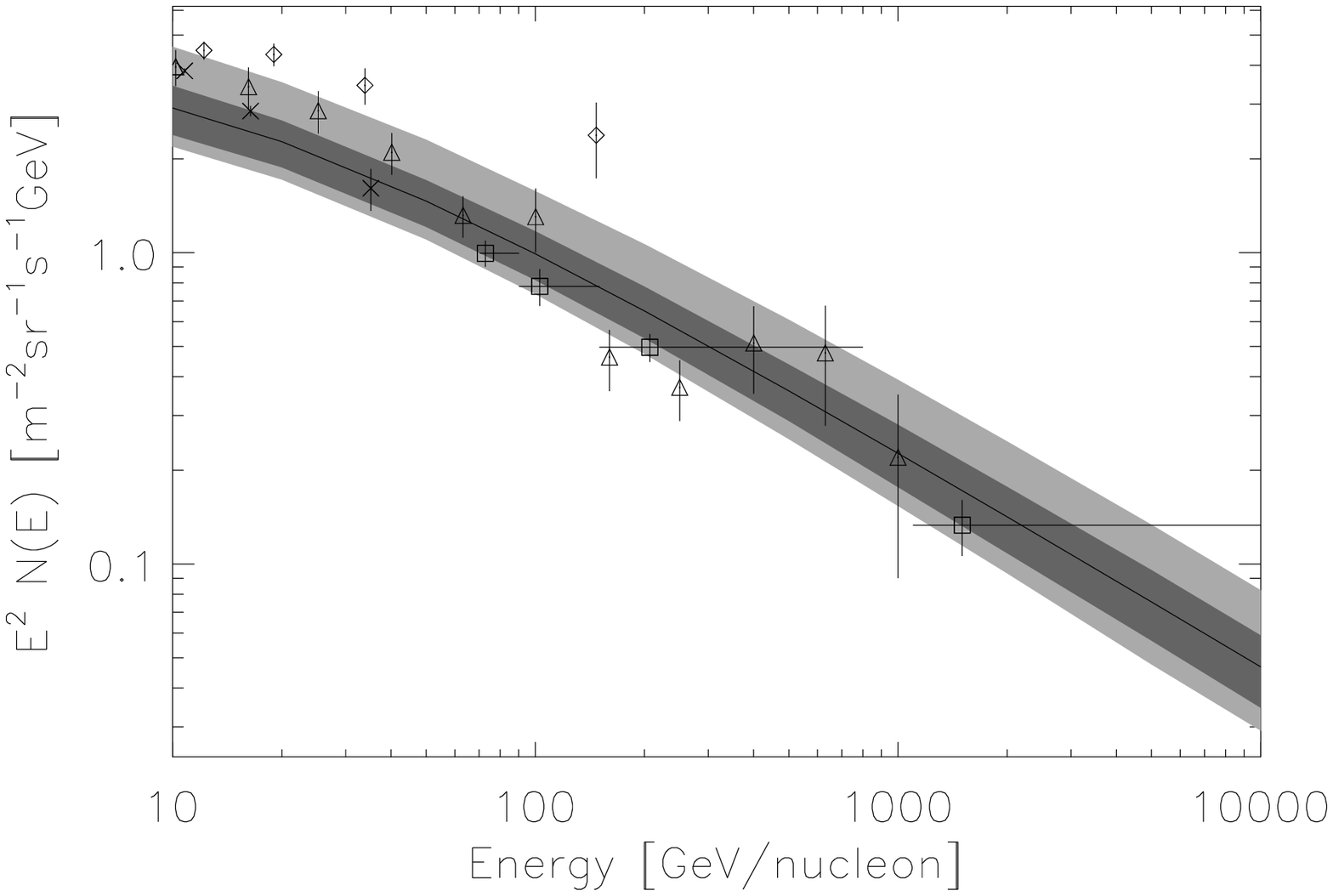}
\vspace{.0pc}
      \end{minipage}
\\
\end{tabular}
\caption{
 The range of pos\-sible $^{16}{\rm O}$ spectra with (lower panel) and without 
(upper panel) Gould's Belt  compared with measurements taken with balloons or satellites.
The dark grey band shows the 68\% containment probability range at each given energy and
the light grey band gives the 95\% range. The solid line marks the averaged (steady-state) 
spectrum. All spectra are as at the top of the atmosphere.  
}
\label{fig:o16specs}
\end{figure}

\begin{figure}
\begin{tabular}{l}
\begin{minipage}{\columnwidth}
\includegraphics[width=0.7\columnwidth]{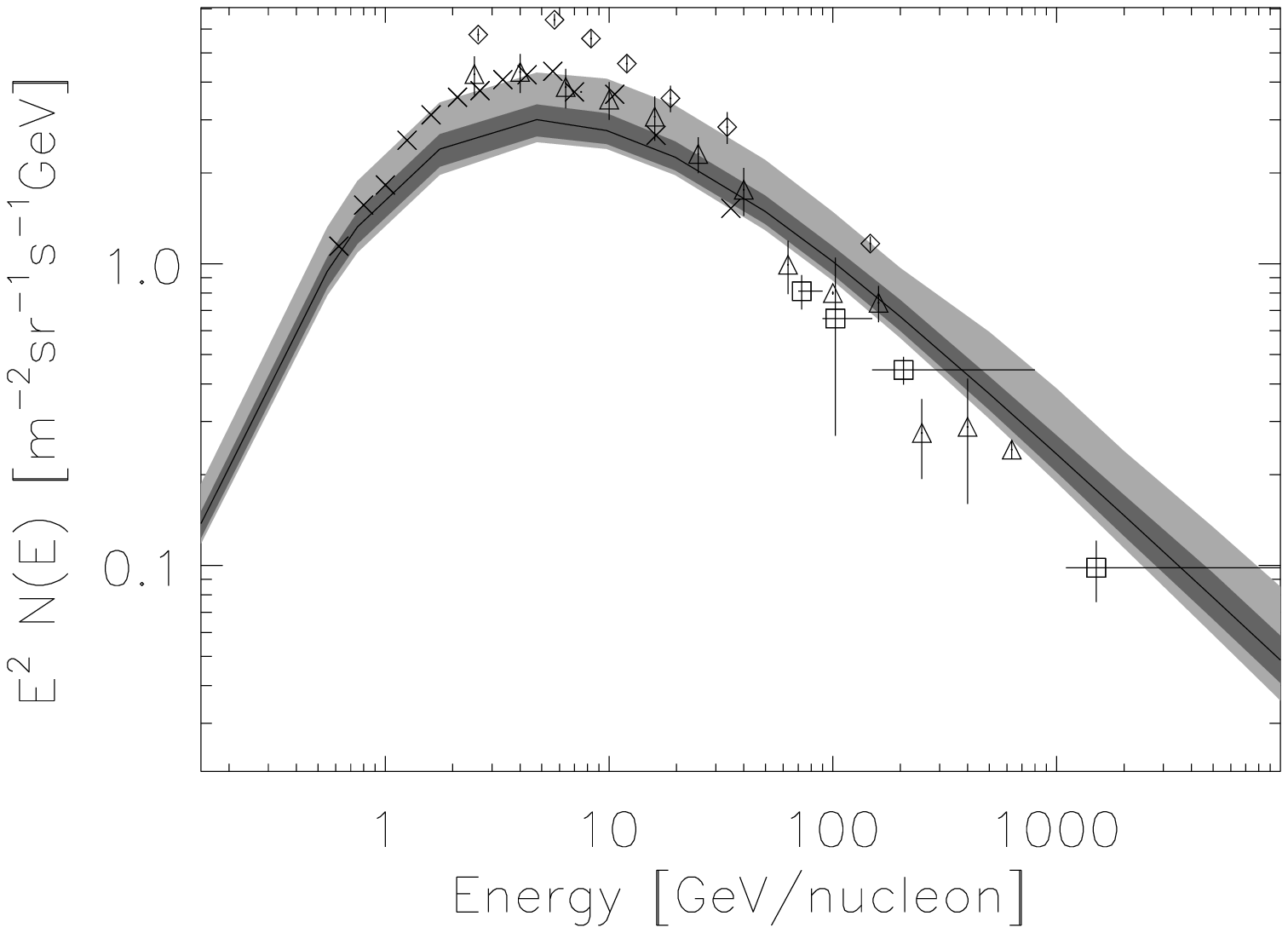}
 \end{minipage}
\\
      \begin{minipage}{\columnwidth}
\includegraphics[width=0.7\columnwidth]{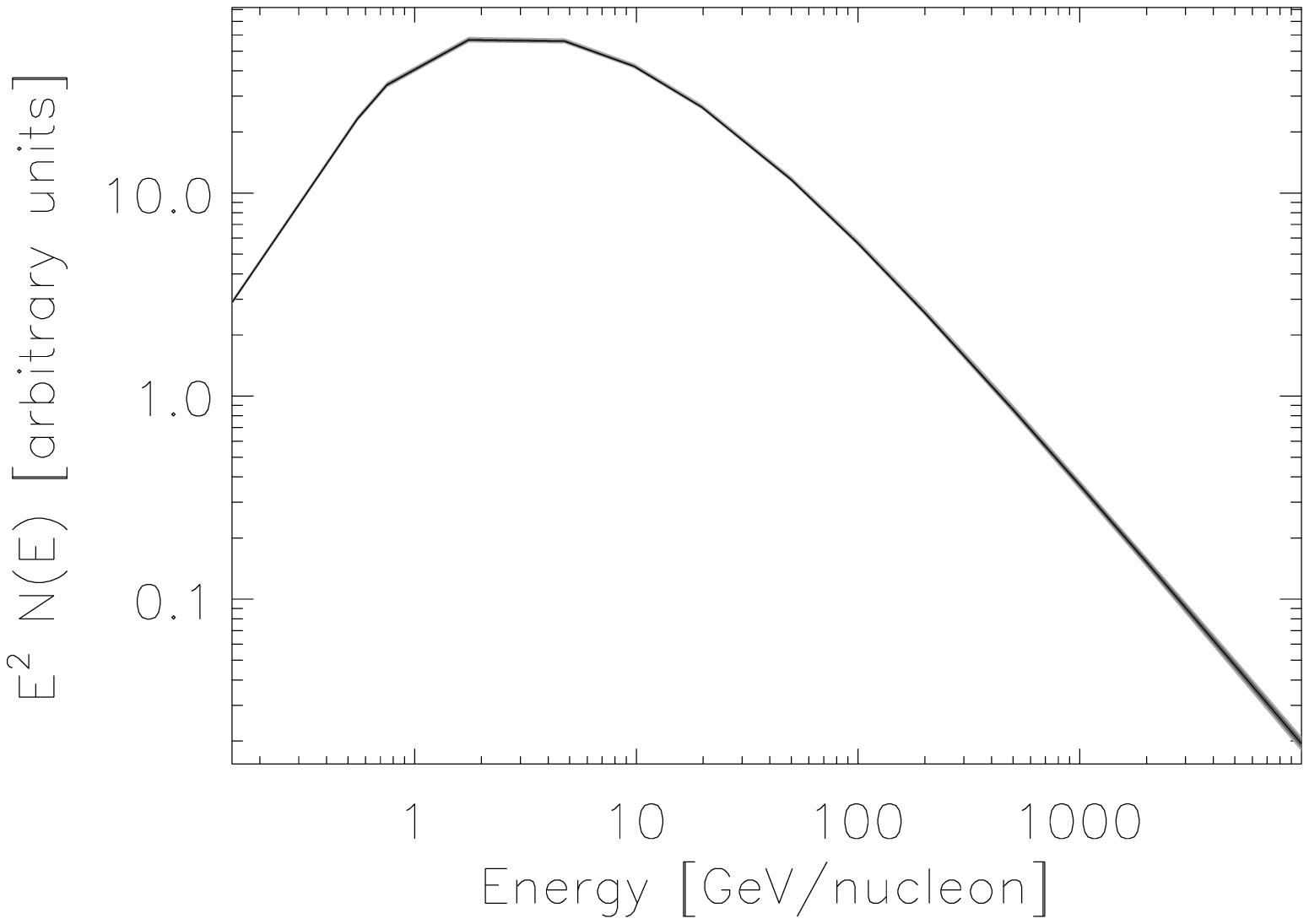}
\vspace{.0pc}
      \end{minipage}
\\
\end{tabular}
\caption{
Same as \Fig{fig:o16specs} but for $^{12}{\rm C}$ (upper panel) and $^{11}{\rm B}$, which is assumed to be 
purely secondary, produced from primary  $^{12}$C only (lower panel). 
}
\label{fig:b11specs}
\end{figure}

In \Fig{fig:o16specs} and \Fig{fig:b11specs} (upper panel), we give the range of possible spectra at the 
position of the Sun for the primary nuclei $^{16}$O and
$^{12}$C, respectively, 
for the case of 
randomly distributed sources (upper panels)
and including also Gould's Belt (lower panels).
In the same figures we compare the range of calculated spectra
with data by \citet{engelmann90} ($\times$), \citet{mueller91} ($\Box$),
\citet{orth78}
($\Diamond$) and \citet{simon80} ($\triangle$). Those authors
provide data 
for  Carbon as well as  oxygen.  
 We model the effect of solar modulation
using the  force field approximation \citep{gleeson68}, assuming a modulation 
parameter of $\Phi\,=\,500$\,MV.
These locally measured data fit quite well in our calculated range of
possible spectra.  

 The amplitude of variations in the possible primary CR nuclei
spectra is clearly seen 
and does not increase much with energy as is the case, e.g., for the electrons 
(section \ref{subse:elektron}). The level of computational noise is very much less than the 
calculated fluctuation amplitudes for the primary CRs, as is indicated by the small level of fluctuations 
calculated for the secondary CRs in section \ref{subse:seko}.
The local clustering of sources (section \ref{sec:clusters}) due to Gould's Belt enhances the
fluctuation amplitude, so the model fits the measurements even better.
We also note a slight steepening of the
averaged spectrum. 

\begin{figure}
\includegraphics[width=\textwidth]{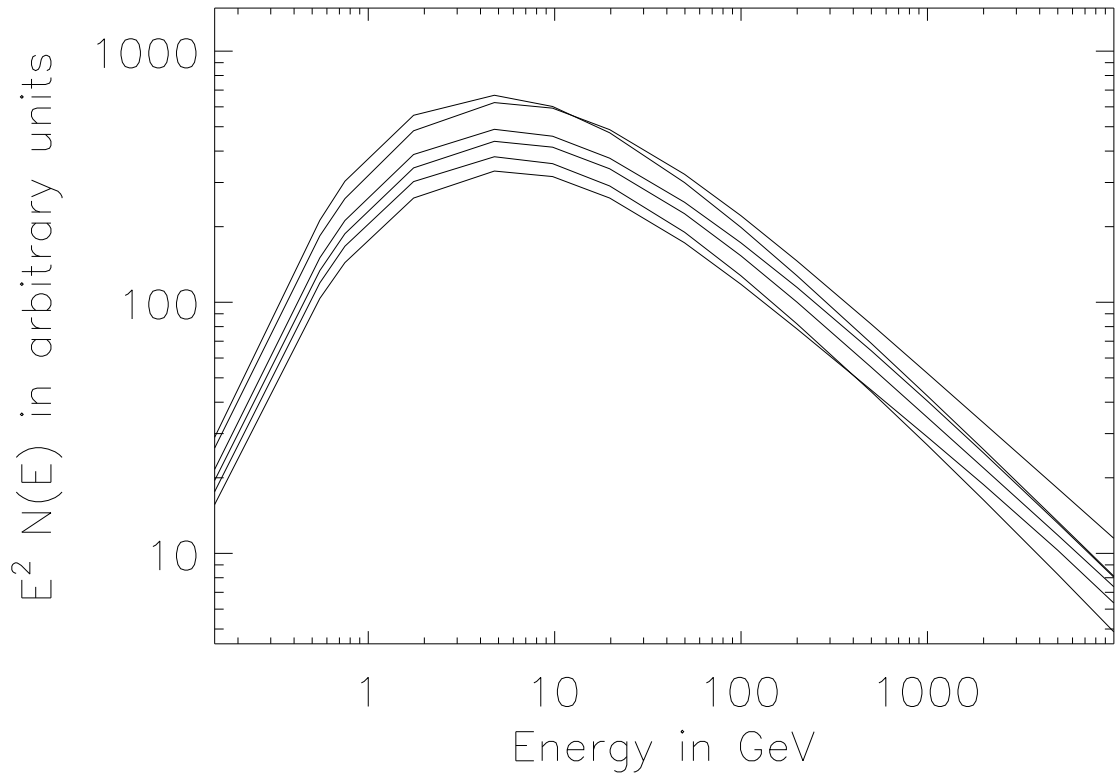}
\caption{Sample of possible $^{12}$C spectra, given at the top of the atmosphere. }
\label{fig:c12bspspec}
\end{figure}

\Fig{fig:c12bspspec} shows some examples of our calculated  $^{12}$C spectra. 
Note that the spectra mostly vary in their amplitudes, and only weakly in their
forms. We now  discuss these findings in detail.

\subsection{Effects due to the Clustering of Sources}
\label{sec:clusters}
We investigated the effects of a local clustering of sources using Gould's Belt as an example.
Our results  are shown in \Fig{fig:o16specs} 
for the CR primary nucleus $^{16}$O and 
in \Fig{fig:b11specs}
for CR secondary nucleus $^{11}$B in each case with and without considering 
Gould's Belt. 
Local clustering of discrete CR sources leads to an enhanced  fluctuation margin in the CR primary spectra
and  on average to a slightly steeper spectrum. 
The slight steepening of the spectrum is  caused by the energy dependence of diffusion: 
the high-energy particles fill a somewhat larger volume than do their low-energy siblings. Then the
differential number density must be softer than the total number spectrum, except for the short time interval
the low-energy particle need to diffusively propagate to the location at which the spectrum is measured.
Almost no effects of Gould's Belt can be  seen in the CR secondary spectra.

\subsection{Implications on the Secondary to Primary Ratio}
\label{subse:seko}
Comparing the range of possible  spectra for  primary CR 
 (see \Fig{fig:o16specs})  and  secondary CR (see \Fig{fig:b11specs}) 
reveals that  the variation in the possible CR nuclei spectra
is much higher for primary nuclei than for secondary nuclei.
Thus, the variation in the primary spectra  should also be  visible in the 
secondary-to-primary ratios. As the  variation in the  amplitude
of the CR primary spectra
is stronger than that of the spectral form, one should expect the 
secondary-to-primary ratios to vary mostly in amplitude but not so much in the 
spectral form. {\citet{taillet03} show using a steady-state
scenario that the CR measured at the earth only probe the 
CR propagation parameters in a rather small domain, and thus outside this
domain the CR fluxes and the secondary-to-primary ratios may
be different. Our calculations
reveal that even if the propagation parameters do not vary, there
is a variation in the secondary-to-primary ratio due to discrete sources.}
 So  provided CR nuclei are accelerated at SNR or other 
kinds of discrete sources, one must account for
fluctuations in the ratio of secondary to
primary CR particles, e.g. the Boron-to-Carbon ratio, which is widely used 
 to determine the parameters of CR propagation models.
In particular, the ratio should decrease in the vicinity of  a source.
Thus it is important to know
whether or not we live in the vicinity of a recent supernova, as has been 
proposed 
by some authors  (see \citet{erlykin97} and subsequent papers).

\subsection{Comparison With Cosmic Ray Electrons}
\label{subse:elektron}
We now  compare our results derived for CR nuclei with 
the findings of \citet{pohl98} for CR electrons.
Evidently, the  fluctuation amplitude in the CR electron spectra strongly grows 
with energy,  whereas that of CR nuclei spectra hardly depends on the particle 
energy (see e.g. \Fig{fig:b11specs}).
Kinks and dents are  possible in CR electron 
spectra but are virtually not seen in the spectra of CR nuclei as shown in Fig.\ref{fig:c12bspspec}.

\section{Summary and discussion}
We  have studied the impact of CR acceleration in SNR
on the spectra of CR nuclei in the Galaxy.
We 
found strong evidence that this assumption leads to CR spectra,  which show
significant variations in
space and time. The behavior of the CR nuclei resembles that of protons, as 
suggested by first computations \citep{strong01},
but differs  considerably  from that of CR electrons \citep{pohl98}.

We have developed a method
to numerically solve the propagation equation based on a series expansion, 
which allows us to use analytical solutions for part of the problem and
an efficient distributed computing for the remainder. 
This method was employed to calculate the CR densities in the Galactic disk with 
high spatial ($\Delta z\,=\, 20\,$pc) and temporal ($\Delta t\,=\,1\,$kyr) resolution for the 
primary  nuclei $^{16}$O, $^{12}$C and
the secondary nucleus $^{11}$B for various distributions of SNR.
This method allowed for the first time to calculate the CR densities in the Galactic disk 
with the 
high spatial and temporal resolution required to follow the CR density fluctuations 
caused by an SNR origin. We also studied the impact of a
locally enhanced SN rate within  
Gould's Belt, a nearby star-forming region.

We found strong variations of the CR nuclei flux by typically 20\% with 
occasional spikes of much higher amplitude,
but only minor changes in the spectral distribution.  
The locally measured primary CR spectra fit well into the  obtained range
of possible spectra. 
We further showed that the spectra of the secondary element Boron 
show almost no variations, so that 
the above  findings also imply significant fluctuations of
the Boron-to-Carbon ratio.
Therefore the commonly used method of determining CR propagation parameters 
by fitting secondary-to-primary ratios appears flawed on account of the
variations that these ratios would show throughout the Galaxy.
 
Some indications that the CR flux varies in the Galaxy are given by the observation
of the diffuse \gr emission in our Galaxy. \citet{digel01}
performed \gr observations of the outer Galaxy.
Their analysis of the diffuse \gr emission from giant molecular clouds
 in the Monoceros region
suggests that the CR flux in the local Galactic arm and the
neighboring Perseus arm differ, i.e. the data suggests an enhancement of
the CR density in the Perseus arm. These observations would be well explained 
by the variations in the flux of CR nuclei that results from a SNR origin.

\citet{hu97} find that the spectrum 
of the diffuse \gr emission toward the inner Galaxy can not be explained by the 
assumption that the locally observed  CR spectrum and electron-to-proton 
ratio hold throughout the Galaxy. At \gr energies above
some GeV, where the spectrum is presumably dominated by CR-nucleon-induced 
radiation, they find that the \gr flux measured exceeds by about 50\%  that
expected if the local CR spectrum and electron-to-proton ratio hold 
throughout the Galaxy. These findings are {
difficult 
to explain by the calculations presented above as being caused by
spectral variations of the CR nuclei flux, for we find these to be rather small.
 We have, however, 
neglected the possibility of a dispersion in the source spectral indices,
which has been proposed as an explanation of} the GeV excess \citep[cf.][]{buesching01}. 

\section{Acknowledgements}
IB  acknowledges support by the Bundesministerium f\"r Bildung und Forschung through DLR  grant 50 OR 0006 and by the Deutsche Forschungsgemeinschaft through Sonderforschungsbereich 591.  
MP acknowledges support by NASA under award No. NAG5-13559.

\appendix
\section{Green's Function for the Spherical Symmetric Propagation Equation}
\label{app:appc}
In case of  spherical symmetry, the propagation equation 
in spherical coordinates reads (\Gl{disc:kugelks}):
\be
\frac{\partial N}{\partial t}
-S
=
k\left(\frac{\partial^2 N}{\partial r^2}+
\frac{2}{r}\frac{\partial N}{\partial r} \right)
-b N.
\label{app:kugelks}
\ee
Using the ansatz 
\be
N&=& {\rm exp}\left(-\frac{1}{T}t\right)\cdot \phi
\label{app:ansatzks1}
\ee
 leads to an equation for $\phi=\phi(r,\varphi, t)$
\be
\frac{\partial \phi}{\partial t}
-{\rm exp}\left(\frac{1}{T}t\right)\cdot S
=
k\left(\frac{\partial^2 \phi}{\partial r^2}+
\frac{2}{r}\frac{\partial \phi}{\partial r} \right).
%-b N
\label{app:kugelksgl2}
\ee
For this equation, a Green's function can be found in the literature \citep{butkovskiy82}. It reads
\be
\bar G&=& \Theta\left(t-t_0\right)\frac{1}{4\pi\sqrt{\pi}\sqrt{k}}\frac{1}{r\,r_0\sqrt{t-t_0}}
\nonumber \\
&&\times {\rm exp}\left(-\frac{r^2+r_0^2}{4\,k\left(t-t_0\right)}\right)
{\rm sinh}\left(\frac{2\,r\,r_0}{4k\left(t-t_0\right)}\right).
\label{app:green1}
\ee
Re-substituting \Gl{app:ansatzks1}, we finally get
\be
G&=& \Theta\left(t-t_0\right)\frac{1}{4\pi\sqrt{\pi}\sqrt{k}}\frac{1}{r\,r_0\sqrt{t-t_0}}
 \exp\left(-b\left(t-t_0\right)\right)
\nonumber \\
&&\times {\rm exp}\left(-\frac{r^2+r_0^2}{4\,k\left(t-t_0\right)}\right)
{\rm sinh}\left(\frac{2\,r\,r_0}{4k\left(t-t_0\right)}\right)
\label{app:greenk}
\ee
which we  now show to be  indeed the desired fundamental  solution of 
\Gl{app:kugelks}.
For the derivatives we have:
\be
\frac{\partial G}{\partial r} &=& -\frac{1}{r}\,G
-
\frac{r}{2k\left(t-t_0\right)}\,G
+
\frac{r_0}{2k\left(t-t_0\right)}
\frac{{\rm cosh}\left(\frac{2\,r\,r_0}{4k\left(t-t_0\right)}\right)}{{\rm sinh}
\left(\frac{2\,r\,r_0}{4k\left(t-t_0\right)}\right)}\,G
\\
\frac{\partial^2 G}{\partial r^2} &=& 
\frac{2}{r^2}\,G 
+
\frac{1}{2k\left(t-t_0\right)}\,G
-
\frac{r_0}{k\,r \left(t-t_0\right)}\frac{{\rm cosh}\left(\frac{2\,r\,r_0}{4k\left(t-t_0\right)}\right)}
{{\rm sinh}\left(\frac{2\,r\,r_0}{4k\left(t-t_0\right)}\right)}\,G 
+
\left(\frac{r}{2k\left(t-t_0\right)}\right)^2G
\nonumber \\
&&-
2\left(\frac{1}{2k\left(t-t_0\right)}\right)^2r\,r_0
\frac{{\rm cosh}\left(\frac{2\,r\,\,r_0}{4k\left(t-t_0\right)}\right)}{{\rm sinh}
\left(\frac{2\,r\,r_0}{4k\left(t-t_0\right)}\right)}G
+
\left(\frac{r_0}{2k\left(t-t_0\right)}\right)^2G
\\
\frac{\partial G}{\partial t} &=&
-
\frac{1}{2\left(t-t_0\right)}\,G
-
b\,G 
+
\frac{r^2+r_0^2}{4k\left(t-t_0\right)^2}\,G
-
\frac{2\,r\,r_0}{4k\left(t-t_0\right)^2} \frac{{\rm cosh}
\left(\frac{2\,r\,\,r_0}{4k\left(t-t_0\right)}\right)}
{{\rm sinh}\left(\frac{2\,r\,\,r_0}{4k\left(t-t_0\right)}\right)}\,G\nonumber\\
&&+\delta\left(t-t_0\right)\frac{1}{\Theta\left(t-t_0\right)}.
\ee
Inserting the above into the \Gl{app:kugelks} yields 
\be
&&\left\{\frac{-1}{2\left(t-t_0\right)}-b +
\frac{r^2+r_0^2}{4k\left(t-t_0\right)^2}
-\frac{2\,r\,r_0}{4k\left(t-t_0\right)^2} 
\frac{{\rm cosh}\left(\frac{2\,r\,\,r_0}{4k\left(t-t_0\right)}\right)}
{{\rm sinh}\left(\frac{2\,r\,\,r_0}{4k\left(t-t_0\right)}\right)}+
\frac{\delta\left(t-t_0\right)}{\Theta\left(t-t_0\right)}\right\}G 
\nonumber \\
&=& k\Biggl\{\frac{2}{r^2}\,G 
+
\frac{1}{2k\left(t-t_0\right)}\,G
-
\frac{r_0}{k\,r \left(t-t_0\right)}\frac{{\rm cosh}\left(\frac{2\,r\,r_0}{4k\left(t-t_0\right)}\right)}
{{\rm sinh}\left(\frac{2\,r\,r_0}{4k\left(t-t_0\right)}\right)}\,G 
+
\left(\frac{r}{2k\left(t-t_0\right)}\right)^2G
\nonumber \\
&&-
2\left(\frac{1}{2k\left(t-t_0\right)}\right)^2r\,r_0
\frac{{\rm cosh}\left(\frac{2\,r\,\,r_0}{4k\left(t-t_0\right)}\right)}
{{\rm sinh}\left(\frac{2\,r\,r_0}{4k\left(t-t_0\right)}\right)}G
+
\left(\frac{r_0}{2k\left(t-t_0\right)}\right)^2G 
\nonumber \\
&&+  \frac{2}{r^2}\,G
-\frac{1}{k\left(t-t_0\right)}\,G
+
\frac{r_0}{r\,k\left(t-t_0\right)}
\frac{{\rm cosh}\left(\frac{2\,r\,r_0}{4k\left(t-t_0\right)}\right)}
{{\rm sinh}\left(\frac{2\,r\,r_0}{4k\left(t-t_0\right)}\right)}\,G
 \Biggr\}
-b\,G +S.
\ee
Comparing terms, we get
\be
&&\left\{\frac{-1}{2\left(t-t_0\right)} 
+\frac{r^2+r_0^2}{4k\left(t-t_0\right)^2}
{{\rm sinh}\left(\frac{2\,r\,\,r_0}{4k\left(t-t_0\right)}\right)}
+\frac{\delta\left(t-t_0\right)}{\Theta\left(t-t_0\right)}\right\}G 
-S \nonumber \\
&=& k\Biggl\{ 
+
\frac{1}{2k\left(t-t_0\right)}\,G
{{\rm sinh}\left(\frac{2\,r\,r_0}{4k\left(t-t_0\right)}\right)}\,G 
+\left(\frac{r}{2k\left(t-t_0\right)}\right)^2G
{{\rm sinh}\left(\frac{2\,r\,r_0}{4k\left(t-t_0\right)}\right)}G
+
\left(\frac{r_0}{2k\left(t-t_0\right)}\right)^2G 
\nonumber \\
&&+  \frac{2}{r^2}\,G
-\frac{1}{k\left(t-t_0\right)}\,G
{{\rm sinh}\left(\frac{2\,r\,r_0}{4k\left(t-t_0\right)}\right)}\,G
 \Biggr\}.
\ee
So finally we have
\be
S&=&\frac{\delta\left(t-t_0\right)}{\Theta\left(t-t_0\right)}G.
\label{app:kugelqd}
\ee
Here $S$ is the delta function for spherical polar coordinates,
so we have to verify that the r.h.s. of \Gl{app:kugelqd} is a representation 
of the delta function 
\be
S&=&\frac{1}{r^2}\delta\left(r-r_0\right)\delta\left(t-t_0\right).
\ee
So with \Gl{app:greenk}, replacing the $\sinh$ by its definition, we have: 
\be
S &=&\delta\left(t-t_0\right)\frac{1}{8\pi\sqrt{\pi}\sqrt{k}}\frac{1}{r\,r_0\sqrt{t-t_0}}
 \exp\left(-b\left(t-t_0\right)\right)
\\ \nonumber
&&\times \left( \exp\left(-\frac{\left(r-r_0\right)^2}{4\,k\left(t-t_0\right)}\right)
-\exp\left(-\frac{\left(r+r_0\right)^2}{4\,k\left(t-t_0\right)}\right)
\right).
\label{app:greenk5}
\ee
Performing the limit for $t\to t_0$ on the r.h.s. leads to
\be
&&\lim_{t\to t_0} \frac{1}{8\pi\sqrt{\pi}\sqrt{k}}\frac{1}{r\,r_0\sqrt{t-t_0}}
 \exp\left(-b\left(t-t_0\right)\right)
\nonumber \\ \label{app:kdeltafk}
&&\times \left( \exp\left(-\frac{\left(r-r_0\right)^2}{4\,k\left(t-t_0\right)}\right)
-\exp\left(-\frac{\left(r+r_0\right)^2}{4\,k\left(t-t_0\right)}\right)
\right) \\
&=& \frac{1}{8\pi\sqrt{\pi}\sqrt{k}} \frac{1}{r\,r_0}\left(
\lim_{t\to t_0}
\frac {\exp\left(-\frac{\left(r-r_0\right)^2}{4\,k\left(t-t_0\right)}\right)}
{\sqrt{t-t_0}}
-\lim_{t\to t_0}
\frac{\exp\left(-\frac{\left(r+r_0\right)^2}{4\,k\left(t-t_0\right)}\right)}
{\sqrt{t-t_0}}
\right)
\nonumber \\
&=&\left\{  { 0 \,\,\,\,\, \,\,\,  \quad {\rm for\,\, } r \not = r_0 
\atop
 \infty \,\,\,\,\,  \quad {\rm for\,\, } r =r_0 } \right. ,
\ee
as  required for a delta function.
To check the normalization of \Gl{app:kdeltafk}, 
we have to integrate over the
spatial domain
\be
I&=&\int_{r=0}^{\infty} \int_{\varphi =0}^{2\pi}\int_{\theta=0}^{\pi} G
r^2\sin \theta dr d\varphi d \theta \Bigg|_{t=t_0}\\
I&=& 4\pi \int_{r=0}^{\infty}  G
r^2 dr\Bigg|_{t=t_0}\\
I&=& \lim_{t_0\to t}\int_0^{\infty}\Biggl\{\frac{r}{8\pi\sqrt{\pi}\sqrt{k}}\frac{1}{r_0\sqrt{t-t_0}}
\nonumber \\
&&\times {\rm exp}\left(-\frac{r^2+r_0^2}{4\,k\left(t-t_0\right)}\right)
{\rm sinh}\left(\frac{2\,r\,r_0}{4k\left(t-t_0\right)}\right)dr\Biggr\}\Bigg|_{t=t_0}
\label{app:greenkd}
\ee
With the integral 3.562.3 of   \citet{gradstein81} 
\be
\int_0^{\infty}r\,\exp\left(-\beta \,r \right)\sinh \left(\gamma \,r \right) dr &=& 
\frac{\gamma}{4\beta}\sqrt{\frac{\pi}{\beta}}\exp\left(\frac{\gamma^2}{4 \beta}\right)
\ee
we finally arrive at
\be
I&=& \lim_{t_0\to t}\Biggl\{ \frac{1}{2\sqrt{\pi}\sqrt{k}}\frac{1}{r_0\sqrt{t-t_0}}
{\rm exp}\left(-\frac{r_0^2}{4\,k\left(t-t_0\right)}\right)
\frac{r_0}{2}\sqrt{\pi\,4\, k \left(t-t_0\right)} 
\nonumber \\
&&\times 
{\rm exp}\left(\frac{r_0^2}{4\,k\left(t-t_0\right)}\right)
\Biggr\}
\label{app:greenkd1}\\
I&=& 1
\ee
which shows that the normalization is correct.

\end{document}